\documentclass[%
 aip,
 amsmath,amssymb,
 reprint,%
]{revtex4-1}

\usepackage{graphicx}
\usepackage{dcolumn}
\usepackage{bm}
\usepackage[english]{babel}
\usepackage[utf8]{inputenc}
\usepackage[T1]{fontenc}
\usepackage{mathptmx}
\usepackage{color}
\usepackage{siunitx}
\usepackage{color}
\usepackage{bm}
\usepackage{multirow}
\usepackage{ifthen}
\usepackage{bbm}
\usepackage{ulem}

\definecolor{orange}{rgb}{1,0.5,0}
\definecolor{darkgreen}{rgb}{0,0.4,0.1}

\DeclareMathOperator*{\argmin}{arg\,min}

\newcommand{\highlightrevision}{false}

\ifthenelse{\equal{\highlightrevision}{true}}
{
\newcommand{\revision}[1]{{\color{red}{#1}}}
\newcommand{\revbar}[1]{{\color{red}{\sout{#1}}}}
}
{
\newcommand{\revision}[1]{{\color{black}{#1}}}
\newcommand{\revbar}[1]{}
}

\begin{document}

\preprint{AIP/123-QED}

\title{Chemi-sorbed versus physi-sorbed surface charge and its impact on electrokinetic transport: carbon versus boron-nitride surface}

\author{Etienne Mangaud}
\affiliation{\small MSME,  Univ Gustave Eiffel, CNRS UMR 8208, Univ Paris Est Creteil, F-77454 Marne-la-Vall\'ee, France}

\author{Marie-Laure Bocquet}
\affiliation{\small PASTEUR, D\'epartement de chimie, \'Ecole normale sup\'erieure, PSL University, Sorbonne Universit\'e, CNRS, 75005 Paris, France}

\author{Lydéric Bocquet}
\affiliation{\small Laboratoire de Physique de l'Ecole normale Supérieure, ENS, Université PSL, CNRS, Sorbonne Universit\'e, Universit\'e de Paris, 75005 Paris, France}

\author{Benjamin Rotenberg}
\affiliation{\small Sorbonne Universit\'e, CNRS, Physicochimie des \'electrolytes et Nanosyst\`emes Interfaciaux, F-75005 Paris, France}
\email{benjamin.rotenberg@sorbonne-universite.fr}

\markboth{}%
{}

\date{\today}


\begin{abstract}
The possibility to control electrokinetic transport through carbon and hexagonal boron nitride (hBN) nanotubes has recently opened new avenues for nanofluidic approaches to face outstanding challenges such as energy production and conversion or water desalination. The $p$H-dependence of experimental transport coefficients point to the sorption of hydroxide ions as the microscopic origin of the surface charge and recent ab initio calculations suggest that these ions behave differently on carbon and hBN, with only physisorption on the former and both physi- and chemisorption on the latter. Using classical non-equilibrium molecular dynamics simulations of interfaces between an aqueous electrolyte and three models of hBN and graphite surfaces, we demonstrate the major influence of the sorption mode of hydroxide ions on the interfacial transport properties. Physisorbed surface charge leads to a considerable enhancement of the surface conductivity as compared to its chemisorbed counterpart, while values of the $\zeta$-potential are less affected . The analysis of the MD results for the surface conductivity and $\zeta$-potential in the framework of Poisson-Boltzmann-Stokes theory, as is usually done to analyze experimental data, further confirms the importance of taking into account both the mobility of surface hydroxide ions and the decrease of the slip length with increasing titratable surface charge density. 
\end{abstract}

\maketitle

\section{Introduction}

The possibility to control the flow of electrolyte solutions through single nanopores and nanotubes has recently opened new avenues for nanofluidic approaches to face outstanding challenges such as energy production and conversion, \textit{e.g.} to harness ``blue energy'' from salinity gradients, or water filtration and desalination~\cite{eijkelNanofluidicsWhatIt2005,noyNanofluidicsCarbonNanotubes2007,siriaNewAvenuesLargescale2017,marbachOsmosisMolecularInsights2019}. Such applications strongly rely on electrokinetic effects, \textit{i.e.} the coupling between water and ionic flows arising at charged interfaces. For decades, the Helmholtz-Smoluchowski theory provided the basic understanding of these couplings on the macroscopic scale, to describe electrokinetic effects in colloidal suspensions or porous media\cite{lyklemaElectrokineticsRelatedPhenomena1995,hunterZetaPotentialColloid2013}. Scaling down to micro- and nanofluidics requires more detailed descriptions of the interface to account for new phenomena, such as fluid slippage on hydrophobic surfaces, ionic correlations and ultimately the discreteness of solvent molecules and ions adopting a layered structure at the surface, in order to control electrokinetic flows and enhance the capabilities of devices~\cite{bocquetNanofluidicsBulkInterfaces2010,pagonabarragaRecentAdvancesModelling2010,rotenbergElectrokineticsInsightsSimulation2013,strioloCarbonWaterInterfaceModeling2016,kavokineFluidsNanoscaleContinuum2021}.

The giant electrokinetic response measured through carbon and hexagonal boron nitride (hBN) nanotubes, which share the same structure but differ in their chemical composition and electronic properties, prompted fundamental questions on the microscopic origin of their surface charge and their different slippage properties\cite{siriaGiantOsmoticEnergy2013,secchiScalingBehaviorIonic2016,secchiMassiveRadiusdependentFlow2016}. The effect of $p$H on this response pointed to the role of hydroxide ions, but the adsorption mechanism remained elusive until recent \textit{ab initio} studies demonstrated that the two surfaces behave rather differently\cite{grosjeanChemisorptionHydroxide2D2016,grosjeanVersatileElectrificationTwodimensional2019,grosjeanSpontaneousLiquidWater2020}. While hydroxide ions are only physisorbed on carbon surfaces, both physisorption and chemisorption have been found on boron nitride, more precisely on the electro-deficient boron atom. Such a difference originates from the different electronic structure of both solids -- semi-metallic for graphene and insulator for hBN -- and has consequences on the mobility of the adsorbed hydroxide anions: chemisorbed ions are fixed on the surface, while physisorbed ones can move along the surface thanks to proton hopping from neighboring solvent molecules. Hence these water self-ions contribute to charge transport and modify the force balance at the interface which controls the slip length. In turn, such hydroxide mobility impacts the electrokinetic response, as shown in recent theoretical studies at the same level of description of Helmholtz-Smoluchowski theory, namely Poisson-Boltzmann theory for the distribution of ions and the Stokes equation (with slip boundary conditions) for the fluid flow~\cite{maduarElectrohydrodynamicsHydrophobicSurfaces2015,silkinaElectroosmoticFlowHydrophobic2019,mouterdeInterfacialTransportMobile2018}.

Experimentally, interfacial properties are inferred from transport coefficients such as the surface conductivity $K_{surf}$ and $\zeta$-potential, which quantify the electric and electro-osmotic current in response to an applied electric field. Direct characterization of the state of the surface under flow can now be achieved on certain surfaces by spectroscopic approaches\cite{oberLiquidFlowReversibly2021}, but to date the surface charge of carbon and boron nitride nanotubes is merely deduced from the electrokinetic measurements via theoretical models. The resulting effective, electrokinetic surface charge density therefore depends on the underlying assumptions and differs from the ``titratable'' surface charge density corresponding to the actual composition of the surface~\cite{bonthuisUnravelingCombinedEffects2012,lyklemaElectrokineticsRelatedPhenomena1995} \revision{(in practice, the charge deduced from an acid-base titration experiment may differ from the actual number of hydroxide ions per unit area, but we will use this term to designate the latter in the following)}. Various refinements of the above-mentioned theory have been proposed, to account \textit{e.g.} for ion specificity~\cite{huangAqueousElectrolytesHydrophobic2008}, the modification of the dielectric properties of water near interfaces~\cite{bonthuisUnravelingCombinedEffects2012} or adsorption/desorption of ions and charge regulation~\cite{zukoskiInterpretationElectrokineticMeasurements1986,manghiRoleChargeRegulation2018,uematsuCrossoverPowerLawExponent2018}. These improvements however come at the price of introducing additional parameters in the description, which raises the questions of their determination from simulations or from experimental data -- and of the ability of these refined models to describe the real interfaces.

Here we investigate the role of the mobility of surface hydroxide ions on the electrokinetic response of carbon and boron nitride surfaces and examine the consequences of this mobility when interpreting the transport coefficients in terms of interfacial properties. To that end, we use all-atoms, non-equilibrium classical molecular dynamics simulations of interfaces between an aqueous solution containing salt with variable concentration and three model surfaces, with chemisorbed or physisorbed hydroxide ions on hBN and with physisorbed hydroxide ions on graphite. These exhaustive simulations, designed with data from previous \textit{ab initio} molecular dynamics \revision{(AIMD)} simulations, provide the surface conductivity and $\zeta$-potential, which we can analyze using the theoretical framework traditionally used to interpret experiments, including in addition the effect of the mobility of physisorbed ions. This analysis provides for each system the effective electrokinetic surface charge density and the slip length, as a function of the titratable surface charge density and of the salt concentration.

The models and methods used to describe the three considered systems are first presented in Section~\ref{sec:SystemMethods}. The molecular simulation results for the surface conductivity $K_{surf}$ and $\zeta$-potential are then reported in Section~\ref{sec:MD}. Section~\ref{sec:analy} introduces the analytical theory used to analyze the transport coefficients, following the approach used to interpret experiments, while Section~\ref{sec:MDtoPB} examines its ability to describe the simulation results. Finally, the corresponding effective electrokinetic surface charge density and slip length are analyzed in Section~\ref{sec:discussion}, which allows to discuss the microscopic origin of the main differences observed between hBN and carbon surfaces and to conclude  the mobility of surface hydroxides when analyzing experimental data.


\section{Systems and methods}
\label{sec:SystemMethods}

\begin{figure}[ht!]
 \includegraphics[width=1.0\columnwidth]{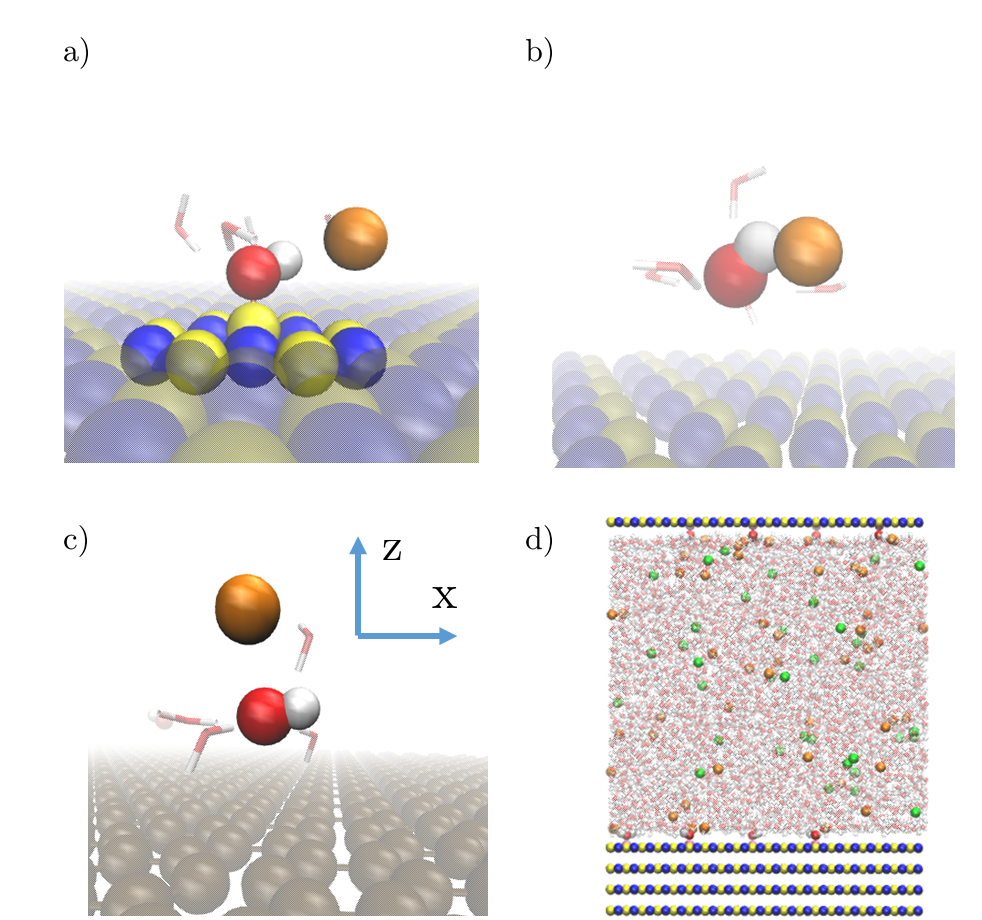}
 \caption{Snapshots of hydroxide defects on hexagonal boron nitride (hBN) and graphite surfaces (G).  (a) chemisorbed hydroxide on hBN, (b) physisorbed hydroxide on hBN, (c) physisorbed hydroxide on G and (d) simulation box in the hBN-C case. Oxygen atoms are in red, hydrogen in white, nitrogen in blue, boron in yellow, carbon in brow, potassium in orange and chloride in green. }
\label{fig:syspres}
\end{figure}

In order to investigate the effect of the mobility of surface hydroxide ions, from which the surface charge originates, we use classical molecular dynamics simulations to study both chemisorbed and physisorbed hydroxide ions on hexagonal boron nitride (hBN) or graphite surfaces (G). Previous experimental~\cite{siriaGiantOsmoticEnergy2013,secchiScalingBehaviorIonic2016} and quantum chemistry\cite{grosjeanChemisorptionHydroxide2D2016,grosjeanVersatileElectrificationTwodimensional2019,al-hamdaniPropertiesWaterBoron2017} studies of hydroxide sorption on these two materials indicate that different mechanism are involved: while HO$^-$ can be either chemisorbed or physisorbed on hBN, precisely with oxygen sitting atop boron atom and hydrogen facing the solution, only physisorption occurs in the case of graphite. Figure~\ref{fig:syspres} illustrates typical configurations of one hydroxide on each type of surfaces as well as the whole simulation box.

Five layers of hBN or graphene are stacked in the $z$ direction and encompass a fluid electrolyte. As three-dimensional periodic boundary conditions are applied, intermediate layers between surfaces avoid spurious interactions between fluid layers close to the surface through the periodic boundary conditions. Simulations are performed in the $NVT$ ensemble, with the equilibrium cell parameters obtained from preliminary simulations in the $NP_zT$ ensemble (at a pressure of $1$~bar), resulting in box sizes of $50.09\times52.06\times65.01$~\AA$^3$ for hBN surfaces and  $51.12\times49.19\times66.92$~\AA$^3$ for G in the x, y and z directions, respectively. 
The electrolyte consists in 4025 water molecules with 7, 36 and 73 pairs of potassium chloride ions (K$^+$ and Cl$^-$), corresponding to salt concentration $c_S$ of 0.1, 0.5 and 1 M. The number of hydroxides (up to 16) is chosen to cover a range of surface charge $\Sigma_{HO^-}$ between 0 and -0.1 C.m$^{-2}$, with additional cations (K$^+$) to ensure the global electroneutrality of the system. Note that in the present work, the hydroxide ions are described classically, and advanced effects like the Grotthus hoping are not considered. These three systems constitute a realistic playground on which we can investigate the response of fixed  (chemisorbed) or mobile (physisorbed) ion to an electric field applied parallel to surface. 

Interatomic interactions are described by 12-6 Lennard-Jones (LJ) pair potential  with Coulombic pairwise interaction and a distance cutoff $r_C$ = 12.0 \AA. Long-distance electrostatic interactions are computed with a particle-particle particle mesh solver (pppm) with a accuracy of 1.10$^{-4}$ relative error on the forces. Water molecules are described with the SPC/E model \cite{berendsenMissingTermEffective1987}. Graphite and hBN surfaces are described by interatomic potentials chosen respectively from ref.~\citenum{werderWaterCarbonInteraction2003} and ref.~\citenum{wonWaterPermeationSubnanometer2007}. The LJ parameters for potassium K$^+$ and chloride Cl$^-$ ions LJ are taken from ref.~\citenum{koneshanSolventStructureDynamics1998}, while those for chemisorbed and physisorbed hydroxides are taken as those of the SPC/E water model.
Lorentz-Berthelot mixing rules are used to compute cross LJ parameters,
except for interactions between surfaces and physisorbed hydroxide ions, where the interaction strength is empirically modified to guarantee that hydroxide ions remain mostly in the vicinity of the surface, in order to match the outcome of ab initio simulations~\cite{grosjeanVersatileElectrificationTwodimensional2019}. Further details, including \revision{initial geometries, force field parameters, and ionic density profiles} are discussed in the Supplementary Material.

Molecular dynamics are carried out with the LAMMPS software package \cite{plimptonFastParallelAlgorithms1995}.  Surfaces are kept fixed throughout all the simulation and water molecules and hydroxides ions are kept rigid (with an OH distance of 1~\AA\ and an HOH angle of 109.47$^\circ$, according to the SPC/E water model) using the SHAKE algorithm \cite{ryckaertNumericalIntegrationCartesian1977} with a tolerance of 1.10$^{-4}$. Whenever relevant, an electric field parallel to the surface of magnitude $E_x = \lVert -\nabla_x V_E \rVert = $ 0.108 V.nm$^{-1}$ is applied in the $x$ direction
\revision{(we have checked that this value is sufficiently small to stay in the linear response regime)}. Trajectories are performed in the $NVT$ ensemble using a timestep of 1~fs and a Nose-Hoover thermostat in the $y$ and $z$ directions with a temperature of 300~K and a relaxation time of 100~fs.
After an equilibration run of 2.5 ns, we perform a production run of 10~ns during which positions and velocities are sampled every 100~fs, from which we compute density profiles and the fluxes discussed in the next section.

\section{Surface conductivity and $\zeta$-potential}
\label{sec:MD}

For each system (hBN and G, chemi- and physisorbed hydroxides respectively noted C and P) and titratable surface charge density $\Sigma_{HO^-}$, defined per unit area of each wall, we sample the \revision{total flux $J$ in the center of the pore (at $z=H/2$)} and charge flux \revision{(averaged over the pore)} $J_q$ 
\begin{equation}
J(\Sigma_{HO^-}) = \revision{\frac{1}{S} \sum_{i=1}^{N_f} v_{x,i} \delta\left(z-\frac{H}{2}\right)} \quad {\rm{and}} \quad J_q(\Sigma_{HO^-}) = \frac{1}{HS} \sum_{i=1}^{N_m} q_i v_{x,i}
\end{equation}
where $S$ is the surface area, $H$ the height of the region occupied by the fluid (estimated from the position of the Gibbs dividing surfaces~\cite{hansenTheorySimpleLiquids2006}), $N_f$ is the total number of fluid particles (water and mobile ions), $N_m$ the number of mobile ions, $v_{x,i}$ velocity of the $i^{th}$ particle in the $x$ direction and $q_i$ its charge. 

The surface conductivity, $K_{surf}$, quantifies the excess electric current arising from the surface charge density $\Sigma_{HO^-}$, in the linear response regime of small applied electric fields $E_x$. We compute it from the charge flux as 
\begin{equation}
K_{surf} (\Sigma_{HO^-}) =  \frac{ \langle J_q (\Sigma_{HO^-})\rangle  - \langle J_q(0) \rangle }{ E_x } \; ,
\label{eq:Ksat}
\end{equation}
where $\langle \cdots \rangle$ denotes the canonical ensemble average of a given observable ans $\langle J_q(0) \rangle$ represents the bulk contribution to the conduction.
The $\zeta-$potential then quantifies the electro-osmotic mobility, resulting from the interfacial charge. Here we  {\it define} the $\zeta$-potential in terms of the total fluid flow according to the Helmholtz expression from the total fluid flow as
\begin{equation}
\zeta(\Sigma_{HO^-}) = - \revision{\frac{  \eta}{\epsilon_r \epsilon_0  E_x \rho_v}} \langle J (\Sigma_{HO^-}) \rangle
\label{eq:zeta}
\end{equation}
where \revision{$\rho_v=N_f/HS$ is the number density of the fluid,} \revision{$\eta=0.729$~mPa.s is the dynamical viscosity of the SPC/E water model at 298~K and 1~bar (from Ref.~\citenum{gonzalezShearViscosityRigid2010})}, $\epsilon_r=70.7$ the relative permittivity of the SPC/E water model \revision{at 298~K and a density of 1~g.cm$^{-3}$} (from Ref.~\citenum{ramireddyDielectricConstantSPC1989}) and $\epsilon_0= 8.85.10^{-12}$~F.m$^{-1}$ the vacuum permittivity. Results are averaged over 9 (hBN-C), 8 (hBN-P)and 6 (G-P) independent trajectories and errorbars are computed from standard errors among trajectories.

 \begin{figure}[ht!]
 \includegraphics[width=1.0\columnwidth]{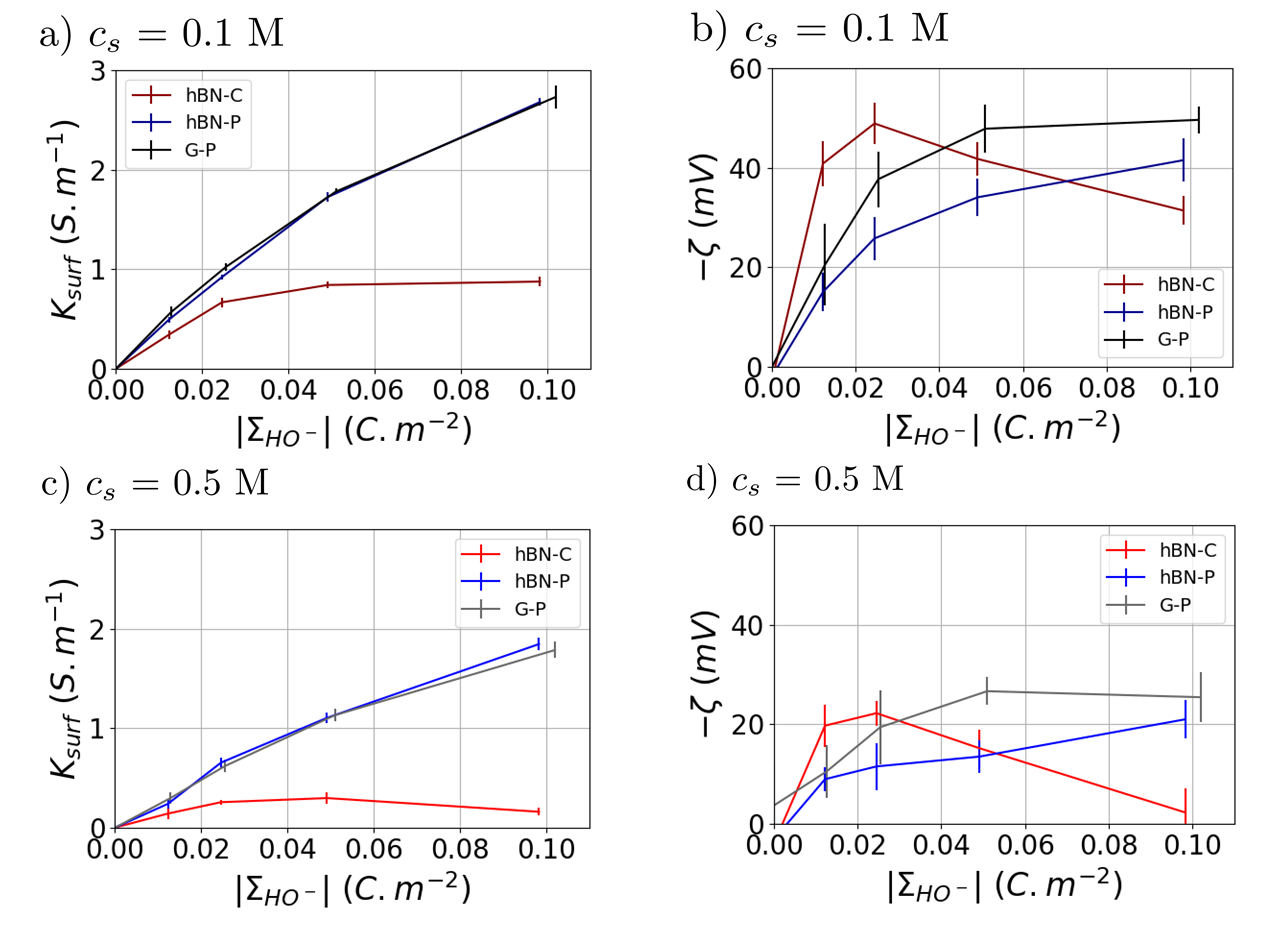}
 \caption{
 Surface conductivity (panels a and c) and $\zeta-$potential (b and d) as a function of the surface charge density $\Sigma_{HO^-}$ of chemisorbed (hBN-C) or physisorbed (hBN-P) ions on hexagonal boron nitride and physisorbed ions on graphite (G-P), for two salt concentrations ($c_S=0.1$~M in panels a and b, $c_S=0.5$~M in panels c and d).}
\label{fig:zetaKsatsimul}
\end{figure}

Figure~\ref{fig:zetaKsatsimul} shows the surface conductivity $K_{surf}$ and $\zeta-$potential as a function of the titratable surface charge density $\Sigma_{HO^-}$ for the three systems (hBN with chemisorbed or physisorbed hydroxides and G with physisorbed hydroxides) and two electrolyte concentrations $c_S=0.1$ and 0.5~M. The results obtained for physisorbed ions on hBN and graphite show very similar trends: $K_{surf}$ increases with $\Sigma_{HO^-}$, while $\zeta-$ initially increases, reaches a maximum near 0.1~C.m$^{-2}$ and then starts decreasing. The agreement between both surfaces is even quantitative for $K_{surf}$. Both physisorbed cases differ from the chemisorbed one, for which $K_{surf}$ is significantly smaller and display a non-monotonic behavior, while $\zeta$ reaches a maximum at a much smaller surface charge density and decays much faster for large surface charge densities. In all cases, both transport coefficients decrease with increasing the salt concentration (see Fig.~\ref{fig:zetaKsatsimul}a versus \ref{fig:zetaKsatsimul}c and \ref{fig:zetaKsatsimul}b versus \ref{fig:zetaKsatsimul}d), as observed experimentally for the $\zeta-$potential~\cite{siriaGiantOsmoticEnergy2013}.

Figure~\ref{fig:zetaKsatsimul} readily exhibits a major effect arising from the soprtion behavior of hydroxide ions : the chemisorbed ions lead to a much reduced surface conductivity by a factor of 3 to 2 depending on the salt concentration. The non-monotonic behavior of the $\zeta$-potential observed in all cases arises from the competition between the larger driving force (more counterions in the fluid) with increasing $\Sigma_{HO^-}$ and  increased friction on the surface walls. In particular, the decrease of $\zeta$ with increasing salt concentration is consistent with a stronger screening of electrostatic interactions (and corresponding thinner electric double layer). 

Since the larger differences are observed between the physisorbed and chemisorbed cases, not between hBN and G with physisorbed ions, the above results shed light on the microscopic origin of the differences observed in experiments with both materials. This however also suggests that the interpretation of experimental data should take this mobility into account. In order to go further, we analyze our results for the transport coefficients using the models used to analyze experiments -- including in addition the mobility of surface hydroxide ions in the physisorbed case. This analysis, together with the ionic charge distribution from molecular simulations, will also provide the basis to interpret the non-monotonic behavior of the $\zeta$-potential (and to a lesser extent of $K_{surf}$ in the chemisorbed case) with the titratable surface charge density $\Sigma_{HO^-}$.

\section{Analytical description}
\label{sec:analy}

The standard theoretical description of electrokinetic phenomena, routinely used to interpret experiments, couples Poisson-Boltzmann (PB) theory for the distribution of (here monovalent) ions and the Stokes equation to describe the flow in the presence of an applied electric field. They are based on a continuous description of the system, with an implicit solvent characterized by its bulk viscosity $\eta$ and dielectric constant $\epsilon_r$, neglecting the finite size of the ions and treating their electrostatic interactions among themselves and with the charged walls at the mean-field level. Extensions have been proposed to improve this descriptions, \textit{e.g.} by introducing permittivity and viscosity profiles in the vicinity of the wall~\cite{bonthuisUnravelingCombinedEffects2012}, or additional mean-field potential to capture ion-specific effects and the influence of the solvent layering near the hard walls~\cite{huangAqueousElectrolytesHydrophobic2008}. We will however restrict ourselves to the simpler version more commonly used in the context of electrokinetic measurements on hBN and carbon nanotubes to predict the surface conductivity and $\zeta$-potential.

In this picture, the two key ingredients to describe the interface are the surface charge density $\Sigma$, which controls the distribution of the ions, and the slip length $b$, which quantifies the balance between viscous stress and friction force and results in a finite velocity of the fluid in contact with the wall (the standard non-slip boundary condition corresponds to $b=0$). Importantly, $\Sigma$ deduced from the experimental transport coefficients is an effective, so-called electrokinetic surface charge density, which may differ from the titratable surface charge density $\Sigma_{HO^-}$ corresponding to the composition of the system.

Two characteristic length scales play an important role in the distribution of the ions at the interface: (a) the Debye length, $\lambda_D= 1/\sqrt{8 \pi l_B c_S}$ with $l_B = e^2/4 \pi \epsilon_0 \epsilon_r k_B T$ the Bjerrum length, $k_B$ the Boltzmann constant and $T$ the temperature, is the length over which electrostatic interactions are screened in the bulk electrolyte, while (b) the Gouy-Chapman length $l_{GC} = e/2\pi  l_B |\Sigma|$ defines the range over which the electrostatic interaction of the monovalent ions with the charged walls dominates the thermal energy. The transport properties are conveniently expressed using the ratio:
\begin{equation}
\chi = \frac{\lambda_D}{l_{GC}} \; .
\end{equation}
The two limits $\chi \ll 1$ and $\chi \gg 1$ correspond to low surface charge / high electrolyte concentrations,  and high surface charge / low concentrations, respectively. Assuming as an initial guess that $\Sigma = \Sigma_{HO^-}$, $\chi$ ranges approximately between 0.1 and 3 depending on the conditions.
Thus, we are dealing with an intermediate regime, where non-linear features cannot be neglected.

The dynamics of the fluid is described at steady-state by the Stokes equation, complemented by a hydrodynamic boundary condition involving a slip length $b$ (which vanishes in the absence of slippage). For physisorbed hydroxide ions, one needs to take into account the mobility of the interfacial ions, which results from their interaction with both the solvent and the wall. As introduced in Ref.~\citenum{mouterdeInterfacialTransportMobile2018}, the corresponding frictions $\lambda_s$ and $\lambda_w$, which enter in the hydroxide mobility as $\mu_{HO^-} =\frac{e}{\lambda_s + \lambda_w}$, in addition to the electrolyte mobility $\mu$. The full expressions of the transport coefficients are~\cite{mouterdeInterfacialTransportMobile2018}:
\begin{align}
K_{surf} = \frac{2}{H} \left[\mu e \vert \Sigma \vert (1+\delta) \frac{\chi}{\sqrt{1+\chi^2} + 1} \right. \nonumber \\ 
\left. + \frac{b}{\eta} (1-\alpha_s)^2 \Sigma^2 + \mu_{HO^-} \vert \Sigma \vert \right] 
\label{eq:Ksatphys}
\end{align}
for the surface conductivity, where $ \delta = 1 / (2 \pi l_B \mu \eta)$ reflects the electro-osmotic contribution to the electric current, and
\begin{equation}
-\zeta = \frac{2 k_B T}{e} \sinh^{-1}(\chi) - (1-\alpha_s)  \Sigma  \frac{b}{\epsilon_0 \epsilon_r} 
\label{eq:zetaphys}
\end{equation}
for the $\zeta$-potential, where in both equations $\alpha_s = \frac{\lambda_s}{\lambda_s + \lambda_w}$. 

The slip length in fact decreases with the surface charge, due to the increased interaction of the fluid with the charged wall~\cite{jolyLiquidFrictionCharged2006}. Using a surface force balance ~\cite{mouterdeInterfacialTransportMobile2018}, the slip length $b$ also depends on the frictions $\lambda_s$ and $\lambda_w$ as
\begin{equation}
b(\Sigma) = \frac{b_0}{1 + \beta_s |\Sigma|/e}
\label{eq:slmod}
\end{equation}
where $b_0$ is the bare slip length and $\beta_s =  \frac{b_0 \lambda_s \lambda_w}{\eta (\lambda_s + \lambda_w)}$. 

The above expressions further allow to recover, in the limit of infinite wall friction on the mobile hydroxide ion, $\lambda_w\to\infty$, simpler expressions applying for the case of chemisorbed hydroxides. Indeed, in this limit we have $\alpha_s\to0$ and $\mu_{HO^-}\to0$ so that Eqs.~\ref{eq:Ksatphys} and~\ref{eq:zetaphys} reduce to:
\begin{equation}
K_{surf} = \frac{2}{H} \left[\mu e \vert \Sigma \vert (1+\delta) \frac{\chi}{\sqrt{1+\chi^2} + 1} +  \frac{b}{\eta}  \Sigma^2 \right] 
\label{eq:Ksatchem}
\end{equation}
and
\begin{equation}
- \zeta = \frac{2 k_B T}{e} \sinh^{-1}(\chi) - \Sigma  \frac{b}{\epsilon_0 \epsilon_r} \; ,
\label{eq:zetachem}
\end{equation}
respectively. In the same limit, and assuming a Stokes relation $\lambda_s=3 \pi \sigma_h \eta$ with an effective hydrodynamic diameter for the chemisorbed ion, Eq.~\ref{eq:slmod} can be rewritten as:
\begin{equation}
b(\Sigma) = \frac{b_0}{1 + 3 \pi \sigma_h b_0 \vert \Sigma \vert/e}
\label{eq:slenhet}
\end{equation}
as proposed in Ref.~\citenum{xieLiquidSolidSlipCharged2020}, where the influence of the charge distribution on the walls was studied in detail. In the other limit, dominated by the solvent friction, $\alpha_s\sim1$ and the terms involving the slip length become less important. 

As mentioned above, the electrokinetic surface charge surface $\Sigma$ entering in these equations may (and in general does) differ from the titratable surface charge density $\Sigma_{HO^-}$. Inspired by the analysis of experiments, we compute $\Sigma$ by determining the value that best fits both transport coefficients, under the constraints that they obey Eqs.~\ref{eq:Ksatphys}, \ref{eq:zetaphys} and~\ref{eq:slmod} (in the physisorbed case) or Eqs.~\ref{eq:Ksatchem}, \ref{eq:zetachem} and~\ref{eq:slenhet} (in the chemisorbed case). Specifically, for each titrable surface charge density $\Sigma_{HO^-}$ and salt concentration $c_S$, we find $\Sigma$ as
\begin{equation}
\Sigma = \argmin_\Sigma \left[ \left(\frac{\Delta K_{surf}}{K_{surf}}\right)^2 + \left(\frac{\Delta \zeta}{\zeta}\right)^2 \right]
\label{eq:sigmaargmin}
\end{equation}
where 
\begin{equation}
\frac{\Delta f}{f} = \frac{f(\Sigma,b(\Sigma)) - f^{MD}(\Sigma_{HO^-})}{f^{MD}(\Sigma_{HO^-})}
\label{eq:sigmaargmin2}
\end{equation}
with $f^{MD}$ the observable from MD simulations, and $f$ computed from the analytical theory. This minimization is performed numerically with L-BFGS-B method (implemented in Scipy libraries~\cite{byrdLimitedMemoryAlgorithm1995}) using as an initial guess of $\Sigma=\Sigma_{HO^-}$ and corresponding $b(\Sigma_{HO^-})$.

This minimization depends on a number of physical quantities. In order to reduce the number of fitting parameters in the analysis of the MD results, some quantities are considered as fixed. For the electrolyte mobility, we use the experimental value $\mu$ = 4.8.10$^{11}$ s.kg$^{-1},$ which is close to that predicted by at 25$^\circ$C and infinite dilution with the forcefield used in the present work~\cite{koneshanSolventStructureDynamics1998} ($\approx$ 4.3.10$^{11}$ s.kg$^{-1}$). In the case of physisorbed (hence mobile) hydroxides, we use $\mu_{OH}$ = 9.10$^{11}$ s.kg$^{-1}$ \revision{that allows to describe the MD results, as shown in Section~\ref{sec:MDtoPB}. This value is of the same order of magnitude yet smaller than a previously reported one (13.10$^{11}$ s.kg$^{-1}$, see the Supplementary Information of Ref.~\citenum{grosjeanVersatileElectrificationTwodimensional2019}), inferred from the experimental bulk diffusion coefficient, \textit{i.e} also taking into account the Grotthus mechanism, which is not captured by the present classical force field.}. The value of $\alpha_s$ is chosen as $\alpha_s=0.8$, in line with the experimental results in Ref.~\citenum{mouterde_molecular_2019}. In addition, the slip length $b_0$ corresponding to the bare hBN surface should be the same for both the physisorbed and chemisorbed cases (the one for graphite is however different). In the following, we show results obtained for given values of $b_0^{hBN}=5.3$~nm and $b_0^G=17.2$~nm (see discussion below) and provide a sensitivity analysis by considering the influence of decreasing or increasing the values by 2 nm. The remaining parameters characterizing the hydrodynamic boundary conditions are $\beta_s$ in Eq.~\ref{eq:slmod} or equivalently the effective hydrodynamic radius $\sigma_h$ in Eq.~\ref{eq:slenhet}. We use $\beta_s^{\chi}=85$~nm$^2$ for the chemisorbed case (hBN-C) and $\beta_s^{\varphi}=48$~nm$^2$ for both physisorbed cases (hBN-P and G-P). The corresponding effective hydrodynamic diameters are 1.7~nm, 9.6~\AA\ and 3.0\AA\ for hBN-C, hBN-P and G-P, respectively.

\section{From molecular to continuous models}
\label{sec:MDtoPB}

\subsection{Chemisorbed hydroxides}

\begin{figure*}[ht!]
 \includegraphics[width=2.\columnwidth]{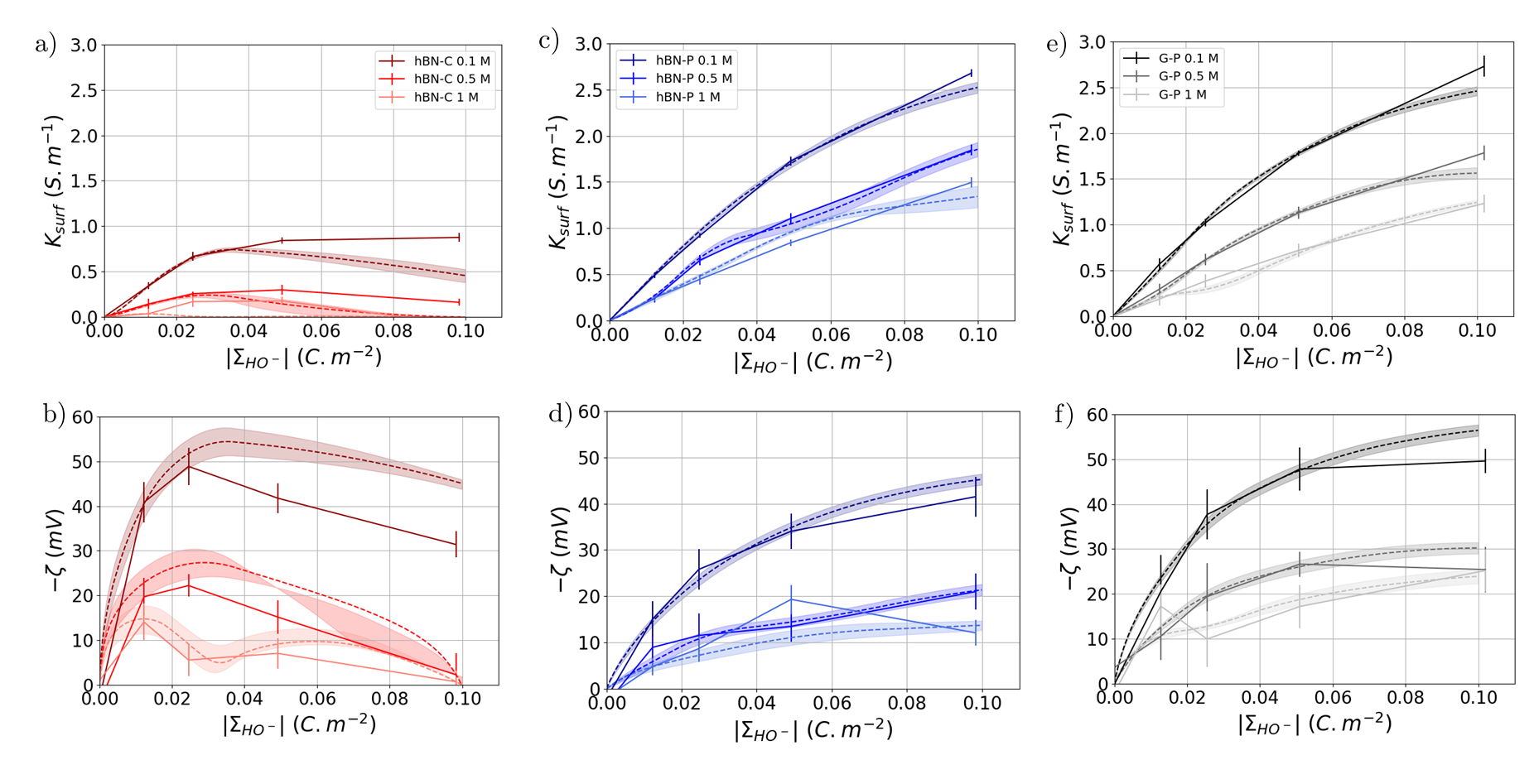}
 \caption{
 Surface conductivity $K_{surf}$ (a, c and e) and $\zeta-$potential (b, d and f) as a function of the titratable surface charge density $\Sigma_{HO^-}$ for hexagonal boron nitride with chemisorbed ions (hBN-C), with physisorbed ions (hBN-P) and for graphite with physisorbed ions (G-P) for three salt concentrations. Solid lines are non-equilibrium MD simulation results (see Eqs.~\ref{eq:Ksat} and~\ref{eq:zeta}), while dashed lines are fits using Eqs.~\ref{eq:Ksatchem} and~\ref{eq:zetachem} for $K_{surf}$ and $\zeta$, respectively (panels a and b) or using Eqs.~\ref{eq:Ksatphys} and~\ref{eq:zetaphys} for $K_{surf}$ and $\zeta$, respectively (panels c, d, e and f). The shaded areas indicate the effect on the predictions of the model of decreasing or increasing the slip length on the bare hBN surface between $b_0^{hBN}$ by 2~nm. 
 }
\label{fig:resfit}
\end{figure*}

We start our analysis of the transport coefficients of Section~\ref{sec:MD} in the framework of the analytical theory of Section~\ref{sec:analy} with the case of chemisorbed hydroxides on hBN. The predictions of Eqs.~\ref{eq:Ksatchem}, \ref{eq:zetachem} and~\ref{eq:slenhet} for $K_{surf}$ and $\zeta$ as a function of the titratable surface charge density $\Sigma_{HO^-}$, using the electrokinetic charges determined from Eqs.~\ref{eq:sigmaargmin} and~\ref{eq:sigmaargmin2} are compared to the simulation results in Fig.~\ref{fig:resfit} (panels (a) and (b)), for three salt concentrations. Even though not quantitative for larger surface charges, the parameterized model correctly captures the evolution of both $K_{surf}$ and $\zeta$ as a function of the titratable surface charge density and the effect of salt concentration. PB theory is not expected to be accurate at concentrations as high as 1~M. Hence, considering all the underlying assumptions, such an agreement is very satisfactory. The slip length for the bare hBN surface, $b_0^{hBN}=5.3$~nm, also used below for the physisorbed hydroxide ions, comparable to previously reported values such as  $3.3\pm0.6$~nm in Ref.~\citenum{tocciFrictionWaterGraphene2014}, 4.0$\pm$0.4~nm in Refs.~\citenum{tocciInitioNanofluidicsDisentangling2020} and~\citenum{poggioli_distinct_2021}. The shaded areas in Fig.~\ref{fig:resfit}, which illustrate the effect of decreasing or increasing $b_0^{hBN}$ by 2~nm, show that the values from the literature would also provide consistent predictions of the transport coefficients.

In Ref.~\citenum{xieLiquidSolidSlipCharged2020}, the authors analyzed the effect of the charge distribution within the solid walls (for a given surface charge density) on the slip length. They obtained Eq.~\ref{eq:slenhet} for heterogenous distributions by considering the friction exerted by the counterions "pinned" by localized charges in a flat wall. In the present case, both the fixed hydroxides and the mobile (though strongly interacting) potassium counterions protrude out of the surface, so that the microscopic interpretation of $\sigma_h$ is less straightforward. We also note that if all counterions were bound to the localized charged sites on the wall, there could be no electro-osmotic flow in the presence of an applied field. Nevertheless, the good agreement of the model with MD results suggests that Eq.~\ref{eq:slenhet} provides a sufficient description of the boundary conditions in Eqs.~\ref{eq:Ksatchem} and~\ref{eq:zetachem} for $K_{surf}$ and $\zeta$, respectively. The large value of the parameter $\sigma_h$ ($\approx1.7$~nm) for this chemisorbed case seems too large to be interpreted as a mere effective hydrodynamic diameter (assuming the validity of Stokes' law and a typical diffusion coefficient for the ions results in an order of magnitude smaller). This is however not surprizing, considering the complexity of the surface with HO$^-$ groups protruding from the surface, around which counterions and water molecules are organized. A larger value of $\sigma_h$ is consistent with the stronger friction resulting from the interactions of the fluid with these interfacial structures.

\subsection{Physisorbed hydroxides}

We now turn to the cases of physisorbed hydroxides on hBN and graphite surfaces. The predictions of Eqs.~\ref{eq:Ksatphys}, \ref{eq:zetaphys} and~\ref{eq:slmod} for $K_{surf}$ and $\zeta$ as a function of the titratable surface charge density $\Sigma_{HO^-}$, using the electrokinetic charges determined from Eqs.~\ref{eq:sigmaargmin} and~\ref{eq:sigmaargmin2} are compared to the simulation results for three salt concentrations in Figs.~\ref{fig:resfit} (panels (c), (d), (e) and (f)) for hBN and G surfaces, respectively. In both cases, model predictions are in even better agreement with the simulation results than for the previous case of chemisorbed hydroxide ions on hBN. Considering again all the underlying assumptions of the model and the fact that we some parameters were not adjusted beyond an initial reasonable guess, such a semi-quantitative agreement is very satisfactory. Here again, deviations are larger for the largest concentration of 1~M were PB theory is not expected to be sufficient, but the trends are still well captured.

The values retained for the slip length, $b_0^{hBN}=5.3$~nm and $b_0^G=17.2$~nm boron nitride and graphite, respectively, are consistent with previous molecular simulation studies. For graphite, reported values cover a rather wide range: $10.6\pm2.2$~nm in Ref.~\citenum{tocciFrictionWaterGraphene2014}, $19.6\pm1.9$~nm in Ref.~\citenum{tocciInitioNanofluidicsDisentangling2020}, $60\pm6$~nm in Ref.~\citenum{kumarkannamSlipLengthWater2012}. Poggioli and Limmer have also reported several estimates for both surfaces, from complementary approaches yielding values around 40~nm for graphite and 6~nm for boron nitride~\cite{poggioli_distinct_2021}. While the range for graphite is quite large, shaded areas in Fig.~\ref{fig:resfit} (panels (c), (d), (e), (f)), which illustrate the effect of decreasing or increasing $b_0^{hBN}$ and $b_0^G$ by 2~nm, show that other values from the literature might also provide reasonable predictions of the transport coefficients..
The values of the parameter $\sigma_h=9.6$~\AA\ and 3.0~\AA\ for hBN-P and G-P, respectively, are smaller than in the chemisorbed case, and more comparable to molecular sizes, so that their interpretation as an effective hydrodynamic diameter seems more justified. This is expected, since the picture of "pinned" counterions due to the localized charge inside the surface (see Ref.~\citenum{xieLiquidSolidSlipCharged2020}) seems more consistent with the case of physisorbed ions. However, the present case remains different, since the walls are overall neutral and the surface charge arises from the adsorption of the hydroxides.
Despite these limitations, we can conclude that the model taking into account the surface mobility faithfully describes the transport coefficients from molecular simulations when applied with reasonable values of the physical parameters.  \revision{We note that this continuous description neglects the change in the mobility of K$^+$ and Cl$^-$ near the surface, which could be influenced by the chemi/physisorbed nature of hydroxide ions, showing that this effect is overwhelmed by the difference in mobility of the latter.}

\subsection{Contributions of slippage and surface mobility}

\begin{figure}[ht!]
 \includegraphics[width=0.8\columnwidth]{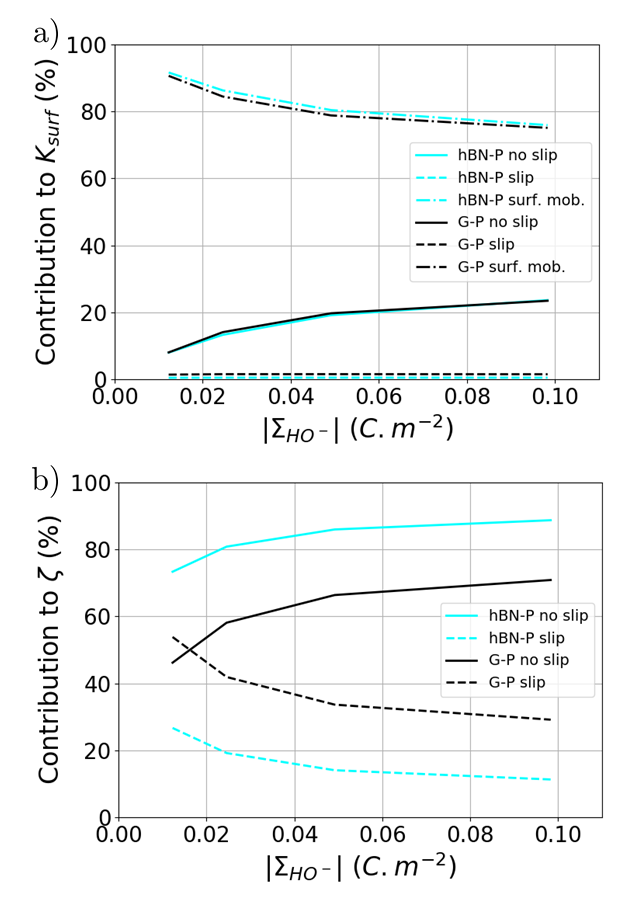}
 \caption{
 Contributions to the surface conductivity $K_{surf}$ (a) and $\zeta-$potential (b) as a function of the titratable surface charge density $\Sigma_{HO^-}$ for hexagonal boron nitride (hBN-P, dark blue lines) and graphite (G-P, black lines) with physisorbed ions for a salt concentration $c_s=0.1$~M. The contributions include the one in the absence of slippage and surface conductivity (solid lines, first term in Eqs.~\ref{eq:Ksatphys} and~\ref{eq:zetaphys}), that of slippage (dashed lines, second term in Eqs.~\ref{eq:Ksatphys} and~\ref{eq:zetaphys}) and that of surface mobility (dashed-dotted lines, third term in Eqs.~\ref{eq:Ksatphys}).
 }
\label{fig:contributions}
\end{figure}

In order to quantify the contributions of slippage and surface mobility to the surface conductivity and $\zeta-$potential in the physisorbed cases, we report in Fig.~\ref{fig:contributions} the fraction of the transport coefficients resulting from the various terms in Eqs.~\ref{eq:Ksatphys} and~\ref{eq:zetaphys}: the first corresponds to no-slip boundary conditions and no surface mobility, the second to the effect of slippage and the third (for $K_{surf}$ only) to the effect of surface mobility. Results are shown for both hBN and G surface with physisorbed hydroxide ions with a salt concentration $c_s=0.1$~M (the trends are similar for other salt concentrations).

It is obvious from Fig.~\ref{fig:contributions}a that the main contribution to $K_{surf}$ over the considered range of titratable surface charge density $\Sigma_{HO^-}$ corresponds to the mobile surface hydroxide ions and that slippage plays a minor role on this transport coefficient, for both hBN and graphite surfaces. This underlines the importance of taking this surface mobility into account to properly describe the molecular simulation results in the case of physisorbed ions. In contrast, this mobility does not contribute to the $\zeta-$potential, but the contribution from slippage is significant. It is more important for graphite, consistently with the larger slip length of the bare graphite surface $b_0^G$ compared to hBN. Despite the decrease of the slip length with increasing titratable surface charge density, this contribution cannot be neglected even for the largest considered $\Sigma_{HO^-}$.


\section{From transport coefficients to effective interfacial properties}
\label{sec:discussion}

\subsection{Electrokinetic surface charge density and slip length}

The results of Section~\ref{sec:MDtoPB} show that is is possible to describe the surface conductivity $K_{surf}$ and the $\zeta$-potential from non-equilibrium MD simulations using an analytical model combining Poisson-Boltzmann theory for the distribution of the ions and the Stokes equation for the solvent dynamics, taking into account slippage of the fluid at the surface and the mobility of surface hydroxide ions when the latter are physisorbed. The effect of the titratable surface charge density $\Sigma_{HO^-}$ and salt concentration $c_s$ can be relatively well captured on the three surfaces, provided that the effective electrokinetic surface charge density $\Sigma$ and the slip length $b$ are adjusted so as to best reproduce the MD results for $K_{surf}$ and $\zeta$ (see Eqs.~\ref{eq:Ksatphys}, \ref{eq:zetaphys} and~\ref{eq:slmod}) -- following the approach used to analyze experimental data. All results so far were presented as a function of the titratable surface charge density $\Sigma_{HO^-}$ and we now examine the electrokinetic surface charge density $\Sigma$ and the slip length $b$ resulting from this procedure. 

\begin{figure}[ht!]
 \includegraphics[width=1.0\columnwidth]{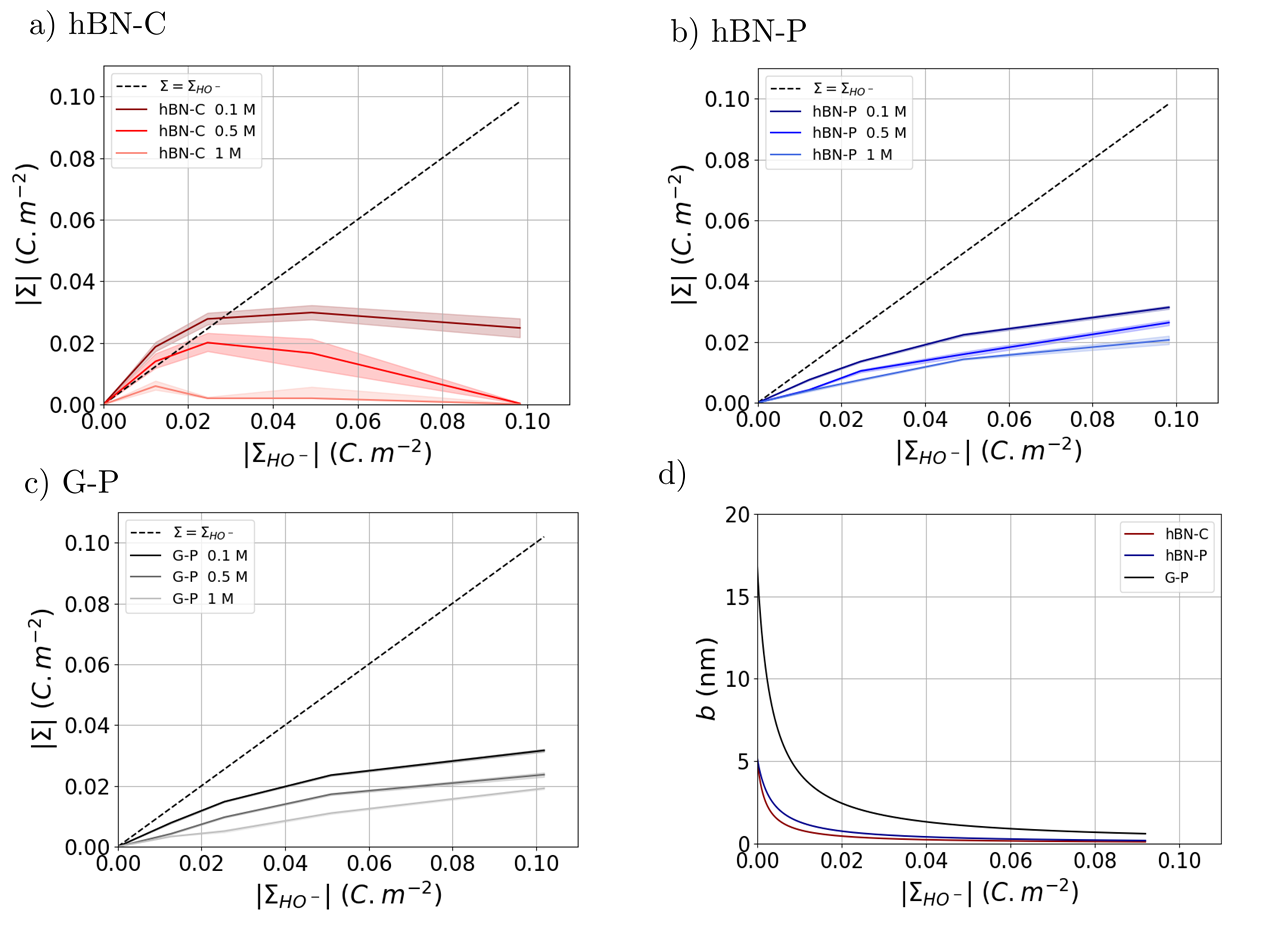}
 \caption{
Electrokinetic surface charge density $\Sigma$ and slip length $b$ obtained from the surface conductivity $K_{surf}$ and $\zeta$-potential at the level of Poisson-Boltzmann-Stokes theory, as a function of the titratable surface charge density $\Sigma_{HO^-}$. The electrokinetic charge density is shown for three salt concentrations for chemisorbed (a) or physisorbed (b) ions on hexagonal boron nitride and physisorbed ions on graphite (c), with dashed lines corresponding to $\Sigma=\Sigma_{HO^-}$. The slip length (d) corresponds to Eq.~\ref{eq:slmod} and~\ref{eq:slenhet} for the chemisorbed and physisorbed cases, respectively.
}
\label{fig:fitparam}
\end{figure}

Figs.~\ref{fig:fitparam}a to~\ref{fig:fitparam}c show the electrokinetic surface charge density as a function of the titratable one, for three salt concentrations and three considered surfaces (hBN-C, hBN-P and G-P). In all cases, $\Sigma$ differs significantly from $\Sigma_{HO^-}$, except in the limit of very small surface charge density in the chemisorbed case, and decreases with increasing salt concentration. The behavior is very similar for both physisorbed cases, with $\Sigma$ smaller than $\Sigma_{HO^-}$ and a tendency to saturate for larger surface charge densities. For the chemisorbed one, the electrokinetic surface charge density starts slightly above $\Sigma_{HO^-}$, then displays a maximum (less pronounced and reached for smaller $\Sigma_{HO^-}$ with increasing salt concentration) before decreasing for large surface charge densities. 

For its part, Fig.~\ref{fig:fitparam}d shows that the three systems evolve from a hydrophobic behavior (large $b$) to a hydrophilic one (small $b$) as $\Sigma_{HO^-}$ increases over the considered range. As previously discussed, slip effects are more pronounced on graphene than on hBN. Furthermore, the slip length observed for hBN is smaller in the chemisorbed case than in the physisorbed case, consistently with the larger friction that the hydroxide groups exert on the fluid when they are fixed on the surface. In all cases, the parametrization predicts a decay of the slip length with increasing surface charge, which reflects the stronger interactions of the fluid with the walls. In turn, slip effects on the $\zeta$-potential diminish, while the contribution of the surface charge to the electrostatic potential and thus the electro-osmotic flows increase. This competition contributes to the saturation of the $\zeta$-potential for high surface charge densities.

The difference between titratable and electrokinetic surface charge densities has already been analyzed in detail by Bonthuis and Netz~\cite{bonthuisUnravelingCombinedEffects2012} by going beyond the bare PB description of the electric double layer. By introducing permittivity and viscosity profiles near flat surfaces as well as non-electrostatic ion-specific adsorption in PB theory, they clarified the effect of surface charge, salt concentration and hydrophobic/philic nature of the surface and were able to explain the main experimental trends. In the present case, the electrokinetic charge does not correspond to such a description of the interface, but rather to the optimal model neglecting these interfacial effects beyond standard PB theory. In addition, as mentioned in Section~\ref{sec:MDtoPB}, the presence of chemisorbed hydroxide protruding from the solid walls or the mobility of adsorbed hydroxide ions may result in effects that are not captured by PB-like theories of the interface, and the charge dependence of the boundary conditions was shown to play an important role on electrokinetic transport in Ref.~\citenum{xieLiquidSolidSlipCharged2020}. Nevertheless, the results shown in Fig.~\ref{fig:fitparam} share common features with the findings of Ref.~\citenum{bonthuisUnravelingCombinedEffects2012}, such as the saturation of $\Sigma$ at large $\Sigma_{HO^-}$, and the fact that the former is generally smaller than the latter. 

\begin{figure}[ht!]
 \includegraphics[width=1.0\columnwidth]{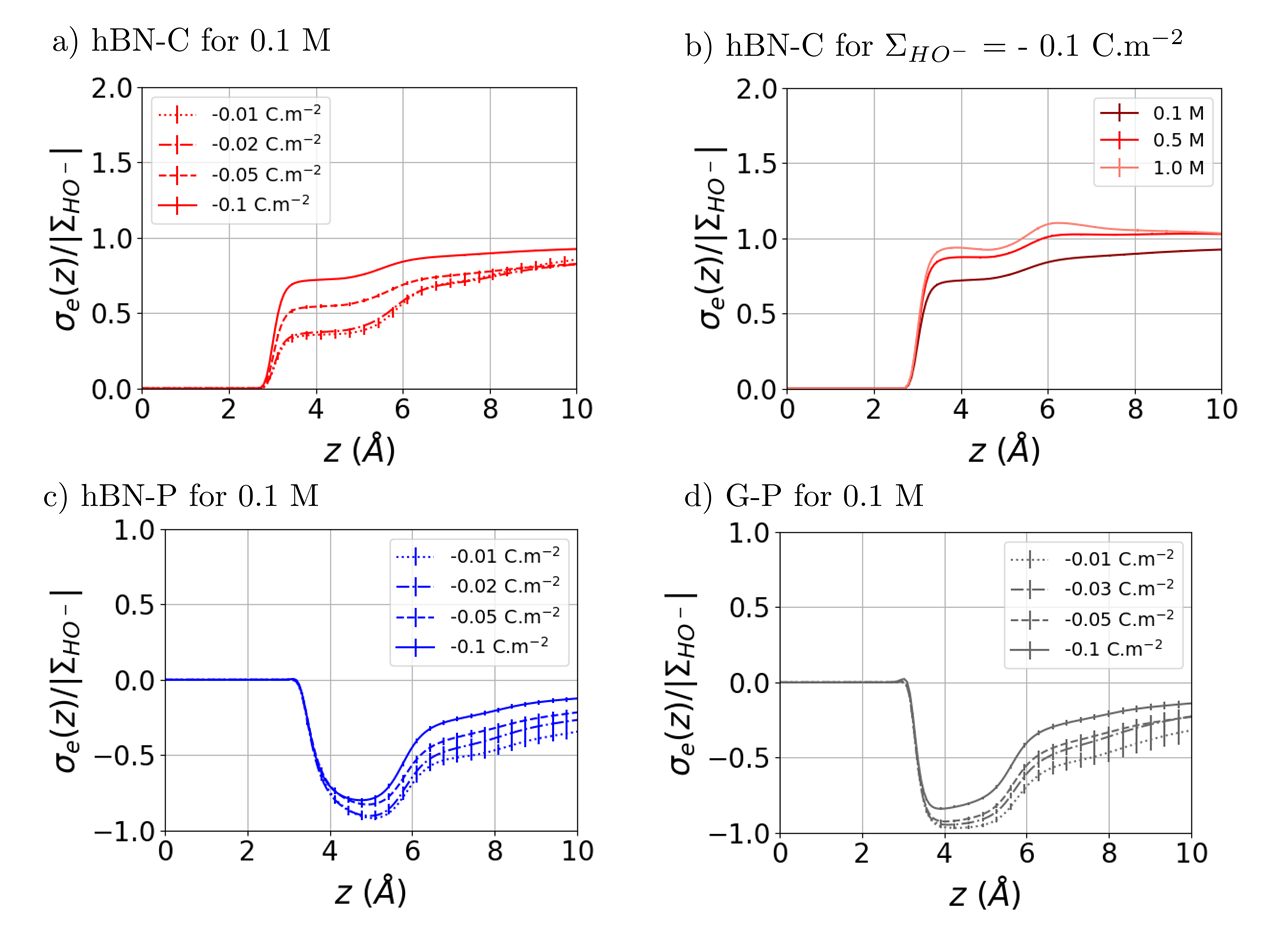}
 \caption{
 Integral of the charge density profile, $\sigma_e(z)$, arising from the ionic distribution at the surface (see Eq.~\ref{eq:iondens}), normalized by the surface charge density $\Sigma_{HO^-}$ (a) for chemisorbed hydroxide ions on hBN and four surface charge densities $\Sigma_{HO^-}$ with fixed salt concentration $c_s=0.1$~M; (b) for the same system as a function of salt concentration with a fixed surface charge density $\Sigma_{HO^-}=-0.1$~C.m$^{-2}$; (c) and (d) for physisorbed hydroxide ions on hBN and G and four surface charge densities $\Sigma_{HO^-}$ with fixed salt concentration $c_s=0.1$~M. In all cases, the position $z=0$~\AA\ corresponds to the first atomic plane of the solid walls.}
\label{fig:densanaly}
\end{figure}

Importantly, the comparison between panels~\ref{fig:fitparam}a and~\ref{fig:fitparam}b shows that for a given composition of the surface (titratable surface charge density $\Sigma_{HO^-}$) the estimate of the electrokinetic surface charge density $\Sigma$ depends on whether or not the mobility of adsorbed hydroxides is taken into account. Conversely, this means that the state of the surface deduced from the transport coefficient might be different from the actual one if one neglects this mobility, as usually done when interpreting experiments. 

\subsection{Charge density profiles}

Since MD simulations give access to the ionic density profiles near the surfaces \revision{(illustrated for the 3 surfaces with a titratable surface charge density $\Sigma_{HO^-}=-0.02$~C.m$^-{2}$ and a salt
concentration $c_s=1.0$~M in Supplementary Materials)}, we analyze the charge distribution at the interface by considering the integrated charge density arising from the mobile ions, defined as
\begin{equation}
\sigma_e(z) = \frac{1}{S} \int_0^z dz' \revision{\left\langle\sum_{i=1}^{N_m} q_i \delta(z'-z_i) \right\rangle}
\label{eq:iondens}
\end{equation}
where $N_m$ is the number of mobile ions (K$^+$, Cl$^-$ and HO$^-$), $q_i$ their charge and $\delta$ the Dirac distribution. These profiles are shown in Fig.~\ref{fig:densanaly} for the three systems, normalized by the $|\Sigma_{HO^-}|$. For chemisorbed hydroxide ions (panels~\ref{fig:densanaly}a and~\ref{fig:densanaly}b), $\sigma_e(z)/|\Sigma_{HO^-}|$ tends to 1 at large distance from the walls, since the total charge of the mobile ions compensates that of the fixed hydroxide sites, while in the physisorbed cases (panels~\ref{fig:densanaly}c and~\ref{fig:densanaly}d) this ratio goes to 0, since all ions are mobile and the fluid is overall neutral. 

The normalization by $|\Sigma_{HO^-}|$ shows that the charge density profiles are only approximately proportional to the titratable surface charge density. While in both physisorbed cases the profiles are (slightly) sub-linear in $|\Sigma_{HO^-}|$ but keep roughly the same shape, in the chemisorbed case the profiles increase more than linearly with $|\Sigma_{HO^-}|>0.02$~C.m$^{-2}$ and change shape with increasing surface charge. One can distinguish two ionic layers, which we describe in the chemisorbed case since they are better separated in that case: a sharper one at $z=3$~\AA, approximately corresponding to cations in the Stern layer (even though here the hydroxide groups protrude from the surface, so that the counterions may be closer to the surface charge than in the standard case), and a wider one between 5 and 7.5~\AA. This second peak flattens and reduces to the benefit of the first as the charge of the surface increases. 

At high surface charges and high salt concentration (see Fig.~\ref{fig:densanaly}b), the charge of the first ionic layers compensates and even exceeds the surface charge arising from the chemisorbed hydroxides. This phenomenon, known as overscreening~\cite{fedorovUnderstandingStructureCapacitance2008,bazantDoubleLayerIonic2011}, contributes to unusual electrokinetic behavior~\cite{rotenbergElectrokineticsInsightsSimulation2013}. In the present case, it may contribute to the non-monotonicity of $K_{surf}$ and the $\zeta$ (see Fig.~\ref{fig:resfit} (panels (a) and (b))), even though other effects such as the charge-dependence of the slip length probably also play a role. In any case, such overscreening is an example of feature not captured by the PB theory used in the present work, which can explain why the transport coefficients from MD simulations are more difficult to describe with the analytical theory for chemisorbed hydroxides. 

No overscreening is observed for the physisorbed cases. The almost identical charge density profiles on hBN and G in that case are consistent with the almost identical electrokinetic charge densities $\Sigma$ (see Figs.~\ref{fig:fitparam}b and~\ref{fig:fitparam}c) and suggests that the main feature leading to the slight differences in $\zeta$-potential between hBN-P and G-P (see Figs.~\ref{fig:resfit}d and~\ref{fig:resfit}f) is the slip length, which is much larger with graphite. Nevertheless, the large difference between the chemisorbed and physisorbed cases for hBN, compared to that between the physisorbed hBN and G cases, supports the conclusion that the main difference observed experimentally between the real systems originates indeed from the different sorption mechanisms.

\section{Conclusion}

Using non-equilibrium molecular dynamics simulations of interfaces between an aqueous solution and three models of hBN and graphite surfaces, we have demonstrated the major influence on the sorption mode of hydroxide ions, from which the surface charge originates, on the interfacial transport properties. The comparison between chemisorbed ions on hBN surfaces with physisorbed ions on hBN and graphite surfaces shows that the main difference on the surface conductivity and the $\zeta$-potential and their evolution with the surface charge and salt concentration is observed between the chemisorbed and physisorbed cases, while the effect of the nature of the wall (hBN vs graphite) in the physisorbed cases is more subtle.

The different sorption mechanisms of hydroxide on hBN and graphite were put forward as an explanation of the dramatic effect of the nature of the walls in electrokinetic experiments with nanotubes and confirmed by quantum chemistry calculations. Our molecular simulations, with models based on microscopic insights from quantum chemistry, allow to disentangle the effect of the mobility of surface hydroxide ions on the macroscopic response to an applied electric field from other physical properties such as the surface charge density and the slip length. The analysis of the MD results for the surface conductivity and $\zeta$-potential in the framework of Poisson-Boltzmann-Stokes theory, as is usually done for the analysis of experimental data, further confirms the importance of taking into account both the mobility of surface hydroxide ions and the decrease of the slip length with increasing the titratable surface charge density, which is in fact unknown in experiments. 

The interfacial structure from MD simulations further confirms the role of the hydroxide adsorption mechanism on the distribution of all ionic charges, hence on the electrokinetic response. It also underlines the (expected) limitations of PB theory to predict the interfacial ionic distribution, especially at high surface charge density and concentrations, in particular because the surface hydroxides protrude from the otherwise atomically smooth solid walls. Nevertheless, the main trends can be captured at this level of theory with a proper parameterization of an effective electrokinetic surface charge density. Our analysis further provides the link between the latter and the titratable surface charge density (\textit{i.e.} the actual composition of the surface). Overall, the present approach bridges the gap between quantum chemistry and experiments via classical MD simulations and PB theory and shows that the surface properties, notably the effective electrokinetic charge, deduced from the transport coefficients may not reflect the actual state of the surface if the effect of the surface mobility of adsorbed ions is not taken into account.

In this analysis, we have limited the number of adjustable parameters for simplicity, which entails some arbitrariness in the choice of some of them, such as the crucial mobility of surface hydroxides. The ability to reproduce the electrokinetic transport coefficients justifies a posteriori these choices, but a more microscopic derivation would be desirable. In addition, the force field was designed to capture the essential difference between chemisorbed and physisorbed ions and we only investigated the two limit cases where only one sorption mode is considered, however on real hBN surfaces both could coexist. Finally, in real water the transport of charge in the presence of hydroxide ions may proceed via the Grotthus mechanism, which is not captured by the present classical MD simulations. \revision{This mechanism is expected to increase the conductivity (both in the bulk and at the surface), but it is difficult to make predictions of its implications for other transport properties, in particular the electro-osmotic response and corresponding $\zeta$-potential, without performing simulations enabling its manifestation, such as AIMD.} While a more quantitative comparison with experimental data would require refining both the microcopic description of the interface and the theoretical analysis of the simulation results, we don't expect the main conclusions of this work, namely the fundamental role of the sorption mechanism on the electrokinetic response and the importance of taking the surface mobility into account to interpret this response in terms of interfacial properties, to be contradicted by such improvements of the microscopic model.

\section*{Acknowledgments}

The authors thank Beno\^it Grosjean for providing the AIMD trajectory that served to design the model used in the present work. The authors acknowledge financial support from the French Agence Nationale de la Recherche (ANR) under Grant No. ANR-17-CE09-0046-02 (NEPTUNE), from the H2020-FETOPEN project NANOPHLOW (Grant No. 766972) and from the European Research Council (ERC) under the European Union's Horizon 2020 research and innovation programme (grant agreement No. 863473, consolidator grant SENSES, and 785911, advanced grant SHADOKS). We are grateful for computational resources provided by CINES (Centre Informatique National de l'Enseignement Sup\'erieur) under grant DARI A0090807364.

\section*{Data Availability Statement}

The data that support the findings of this study are available from the corresponding author upon reasonable request.

\end{document}


\preprint{}

\title{Supplementary Material for  ``Chemi-sorbed versus physi-sorbed surface charge and its impact on electrokinetic transport: carbon versus boron-nitride surface''}

\author{Etienne Mangaud}
\affiliation{\small MSME,  Univ Gustave Eiffel, CNRS UMR 8208, Univ Paris Est Creteil, F-77454 Marne-la-Vall\'ee, France}

\author{Marie-Laure Bocquet}
\affiliation{\small PASTEUR, D\'epartement de chimie, \'Ecole normale sup\'erieure, PSL University, Sorbonne Universit\'e, CNRS, 75005 Paris, France}

\author{Lydéric Bocquet}
\affiliation{\small Laboratoire de Physique de l'Ecole normale Supérieure, ENS, Université PSL, CNRS, Sorbonne Universit\'e, Universit\'e de Paris, 75005 Paris, France}

\author{Benjamin Rotenberg}
\affiliation{\small Sorbonne Universit\'e, CNRS, Physicochimie des \'electrolytes et Nanosyst\`emes Interfaciaux, F-75005 Paris, France}
\email{benjamin.rotenberg@sorbonne-universite.fr}

\date{December 2021}

\maketitle

This Supplementary Material presents the model used to describe the interfaces of graphite and hexagonal boron nitride with aqueous solutions in the presence of chemisorbed or physisorbed hydroxyle ions. Section~\ref{sec:geometry} and~\ref{sec:forcefield}  report all necessary details regarding the geometry of the surfaces and the force field, respectively.

\section{Surface geometries}
\label{sec:geometry}

During this study, three types of surfaces have been considered : graphite (GR), hexagonal boron nitride (hBN) and hexagonal boron nitride with hydroxide defects (hBN-d). The data provided in this section correspond to a 2D unit cell which can be periodically extended in the $x$ and $y$ directions. We used $12\times10$ and $5\times6$ cells for a surface cell size of $51.12\times49.19$~\AA$^2$ and $50.09\times52.06$~\AA$^2$ for graphite and hBN surfaces, respectively. 
  
\subsection{Graphite surface}
Graphite surfaces are built with a C-C intralayer bond distance of 1.42~\AA~ and an interplane distance of 3.5~\AA\ (each unit cell contains five planes, translated by (0.71~\AA, 1.23~\AA)  two by two with respect to each other).
We provide the coordinates of the unit cell which reads in XYZ format: 
\small{
\begin{verbatim}
8
4.26  4.91902428  0.0000000
C    0.00000000   0.00000000   0.00000000
C    1.42000000   0.00000000   0.00000000
C    2.13000000   1.22975607   0.00000000
C    3.55000000   1.22975607   0.00000000
C    0.00000000   2.45951215   0.00000000
C    1.42000000   2.45951215   0.00000000
C    2.13000000   3.68926822   0.00000000
C    3.55000000   3.68926822   0.00000000
\end{verbatim}
}

\subsection{Hexagonal boron nitride surface (hBN)}
For hBN surfaces we use the geometry of Ref.~\citenum{wuHexagonalBoronNitride2016}, which reads in XYZ format:
\small{
\begin{verbatim}
32
10.0181   8.676   0
B  -1.2522725   0.723001446   0
B  1.2522725    0.723001446   0
B  0           -1.445998554   0
B  2.504545    -1.445998554   0
N  -1.2522725  -0.723001446   0
N  1.2522725   -0.723001446   0
N  0            1.445998554   0
N  2.504545     1.445998554   0
B  3.7568175    0.723001446   0
B  6.2613625    0.723001446   0
B  5.00909     -1.445998554   0
B  7.513635    -1.445998554   0
N  3.7568175   -0.723001446   0
N  6.2613625   -0.723001446   0
N  5.00909      1.445998554   0
N  7.513635     1.445998554   0
B  -1.2522725   5.061001446   0
B  1.2522725    5.061001446   0
B  0            2.892001446   0
B  2.504545     2.892001446   0
N  -1.2522725   3.614998554   0
N  1.2522725    3.614998554   0
N  0            5.783998554   0
N  2.504545     5.783998554   0
B  3.7568175    5.061001446   0
B  6.2613625    5.061001446   0
B  5.00909      2.892001446   0
B  7.513635     2.892001446   0
N  3.7568175    3.614998554   0
N  6.2613625    3.614998554   0
N  5.00909      5.783998554   0
N  7.513635     5.783998554   0
\end{verbatim}
}
The distance between two successive planes is 3.33 ~\AA~ and, as in the graphite case, a unit cell contains five planes, translated by (0.~\AA, 1.446~\AA) two by two with respect to each other.

\subsection{Hexagonal boron nitride surface with chemisorbed hydroxide (hBN-d)}

We use the same hBN surface geometry has been used except for the hydroxide defects. For each defect, 7 boron atoms are considered (see Section~\ref{sec:forcefield} for the corresponding change in the interfactions): the boron atom to which the hydroxide group is bound, labeled $B^*$ in the following, and its 6 nearest boron neighbors, labeled $B_1$ in the following. 

\begin{figure}[h]
\centering
  \begin{minipage}[c]{.48\linewidth}
        \centering
        \includegraphics[width=1.0\columnwidth]{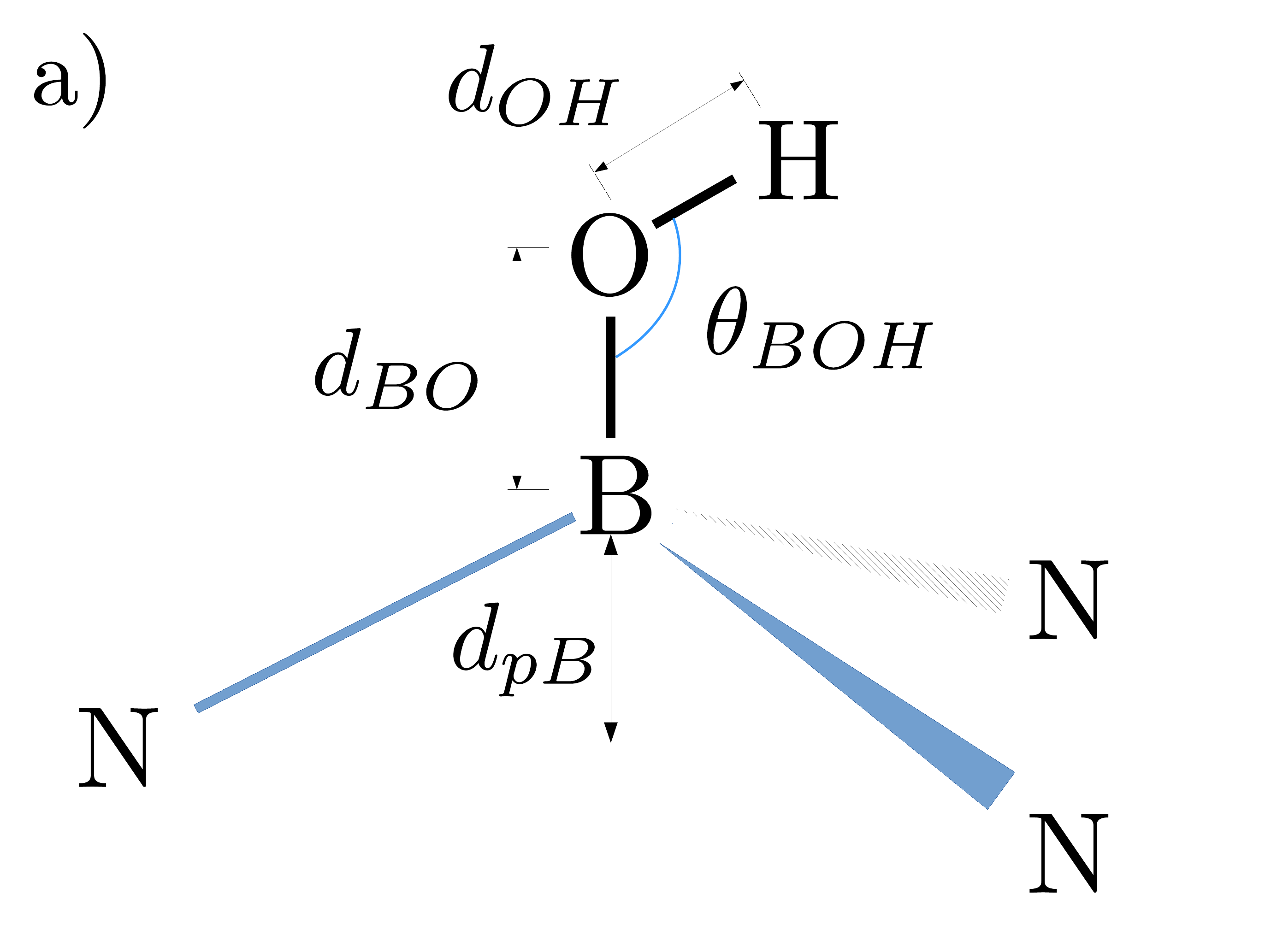}
    \end{minipage}
    \hfill%
    \begin{minipage}[c]{.48\linewidth}
        \centering
        \includegraphics[width=1.0\columnwidth]{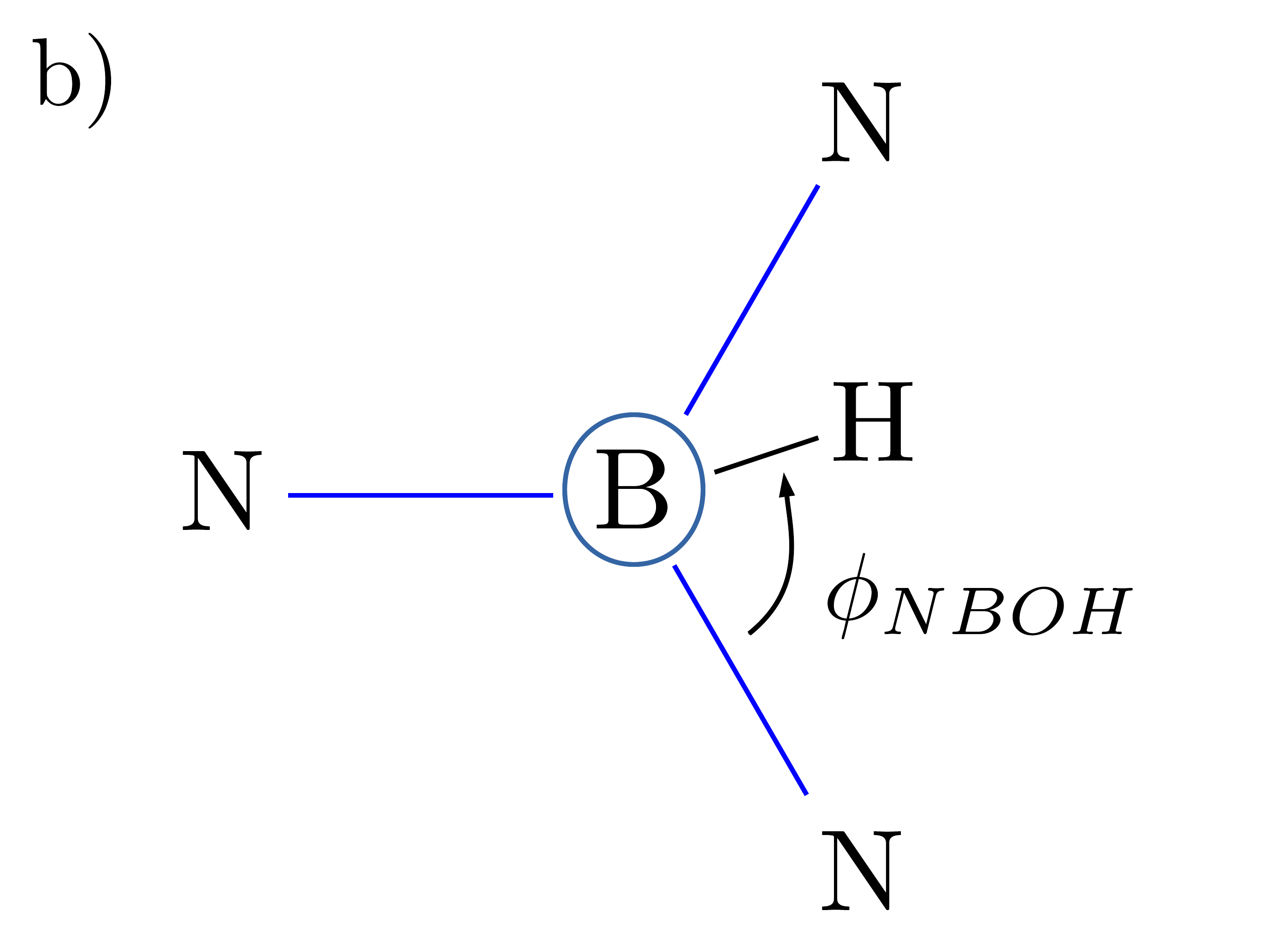}
    \end{minipage}

\caption{(a) Geometry of an OH defect on hexagonal boron nitride. Boron is sp$^3$-hybridized and slightly above the plane defined by three nitrogen's atoms. $d_{OH}$ is the hydroxide bond length, $d_{BO}$ the boron-oxygen bond length, $d_{pB}$ the distance between the nitrogens' barycentre and the boron atom $B^*$ and  $\theta_{BOH}$ the angle between B-O and O-H bonds. (b) Top view illustrating the dihedral angle $\phi_{NBOH}$ between N, B, O and H.}
\label{fig:geom}
\end{figure}

The present model is built using the results obtained by Grosjean \textit{et al.} \cite{grosjeanVersatileElectrificationTwodimensional2019,grosjeanChemisorptionHydroxide2D2016}, by analyzing a 31.145~ps AIMD trajectory carried out by these authors with the CP2K code at the DFT level, with PBE-D3 functional, a DZVP-MOLOPT-SR-GTH basis sets with a plane waves' truncation at 600~Ry energy cutoff, and Geodecker-Teter-Hutter pseudopotentials. Their system consists in a surface made of 30 boron and nitrogen atoms, 97 water molecules and one hydroxide ion chemisorbed on the hBN surface. Fig.~\ref{fig:geom} illustrates the definition of the relevant distances and angles: $d_{OH}$, the distance between the hydroxide's hydrogen and oxygen atoms, $d_{pB}$ distance between $B^*$ and the surface plane, $d_{BO}$ between $B^*$ and $O$ the hydroxide's oxygen, $\theta_{BOH}$ the angle between $B^*$, $O$ and $H$, and $\phi_{NBOH}$ the dihedral angle between the OH and $B^*N$ bonds. 
 
\begin{figure}[h]
\centering

\includegraphics[width=1.0\columnwidth]{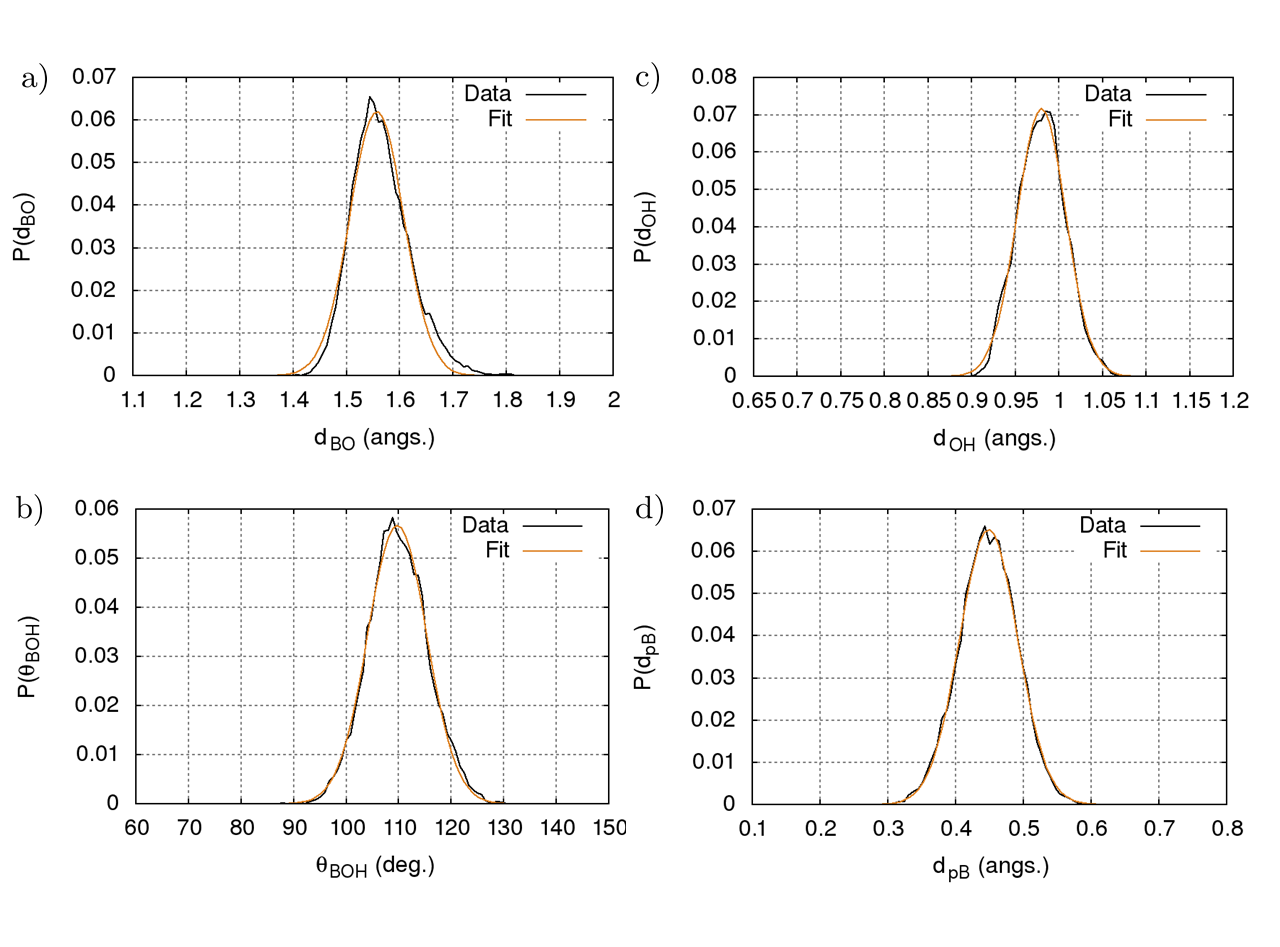}

\label{fig:distrdefect}
\caption{Distributions of the hydroxide bond length $d_{OH}$ (top left), the boron-oxygen bond length $d_{BO}$ (top right), the plan-boron distance $d_{pB}$ (bottom left) and the boron-oxygen-hydrogen angle $\theta_{BOH}$.
}
\end{figure}

 In DFT simulations either in vacuum or in water, the boron site on which the hydroxide is chemisorbed is hybridized in sp$^3$. This results in a tetrahedral configuration with $B^*$ above the plane defined with the three nearest nitrogen atoms. Here, we neglect any other deformation of the surface. We sample from the DFT trajectory the distributions of $d_{OH}$, $d_{BO}$, $d_{pB}$ and  $\theta_{BOH}$ and fit them with Gaussian functions:
 \begin{equation}
 f_x(x) = A_x e^{-\frac{(x-\mu_x)^2}{\sigma_x^2}}
 \, ,
 \end{equation}
 with $x\in\{d_{OH}, d_{BO}, d_{pB}, \theta_{BOH}\}$ to obtain in particular the average value $\mu_x$ for all these properties. The distributions and their fits are shown in Fig.~\ref{fig:distrdefect} and the corresponding averages are summarized in Table~\ref{tab:param}, which also includes for comparison the values reported in vacuum by Grosjean \textit{et al.}~\cite{grosjeanVersatileElectrificationTwodimensional2019, grosjeanChemisorptionHydroxide2D2016}. For all these parameters, the fit performs well. In the presence of water, the distances $d_{OH}$ and $d_{BO}$ and the angle $\theta_{BOH}$ are close to those in vacuum but the distance $d_{pB}$ is reduced by approximately 35\%, suggesting that hybridization is less pronounced in that case.

\begin{table}
\begin{tabular}{|c|c|c|}
\hline 
 & Vacuum & With water \\ 
\hline 
$\mu_{d_{OH}} \pm \sigma_{d_{OH}}$ (\AA) & 0.97 & 0.98 $\pm$ 0.04 \\ 
\hline 
$\mu_{d_{BO}} \pm \sigma_{d_{BO}}$ (\AA) & 1.50 & 1.56 $\pm$ 0.07 \\ 
\hline 
$\mu_{d_{pB}} \pm \sigma_{d_{pB}}$ (\AA) & 0.7 & 0.45  $\pm$ 0.06\\ 
\hline 
$\mu_{\theta_{BOH}} \pm \sigma_{\theta_{BOH}}$ (°) & 109 & 109.76 $\pm$  8.0\\ 
\hline 
\end{tabular}
\label{tab:param}
\caption{
Average bond lengths and angles (see Fig.~\ref{fig:geom}) obtained from DFT calculations in vacuum~\cite{grosjeanVersatileElectrificationTwodimensional2019,grosjeanChemisorptionHydroxide2D2016} and in water (see text for details).
}
\end{table}

\begin{figure}[ht!]
\centering
\includegraphics[width=1.0\columnwidth]{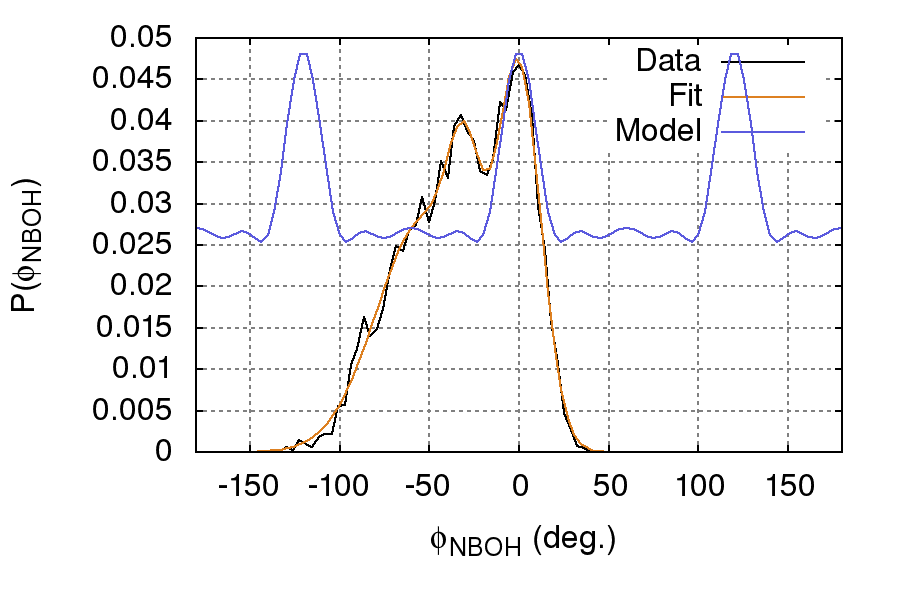}
\caption{Distribution of the dihedral angle $\phi_{NBOH}$. Results sampled from the AIMD trajectory (black line) are compared with the fit by Eq.~\ref{eq:fitphi} (orange line). The blue line is the resulting model distribution, if the AIMD simulations were sufficiently long to overcome possible large energy barriers.
}
\label{fig:phidistr}
\end{figure}

The distribution of the orientation of the chemisorbed hydroxide group with respect to the surface is more complex. We have computed both the dihedral angle between N, B, O and H and the projection of the OH on the nitrogens' plane, which yield similar results. Fig.~\ref{fig:phidistr} shows the distribution of the angle $\phi_{NBOH}$, with respect to a reference $\phi_{NBOH,k} =0$ corresponding to a O-H bond oriented along a B-N bond. The distribution is rather broad and is fitted by a sum of Gaussians:
\begin{equation}
 g(\phi_{NBOH}) = \sum_{k=1}^3 A_k e^{-\frac{(\phi_{NBOH}-\mu_k)^2}{\sigma_k^2}}
 \label{eq:fitphi}
 \end{equation}
which is also shown (orange line) in Fig.~\ref{fig:phidistr}. The parameters corresponding to this fit are summarized in Table~\ref{tab:phiparam}.

\begin{table}[ht!]
\begin{tabular}{|c|c|c|c|}
\hline 
$k$ & 1 & 2 & 3  \\ 
\hline 
$\mu_k$ (°) &  0.14 & -30.24 & -53.23  \\ 
\hline 
$\sigma_k$ (°) & 17.4 & 37.15 & 13.0 \\
\hline
$A_k $ &  0.04 & 0.03 & 0.02 \\
\hline
\end{tabular}
\caption{Expected value $\mu_k$, width $\sigma_k$ and weight $A_k$ of three Gaussian functions in fitting function $g(x)$ (see Eq.~\ref{eq:fitphi}) for the distribution of the dihedral angle $\phi_{NBOH}$ (see Fig~.\ref{fig:phidistr}).
}
\label{tab:phiparam}
\end{table}

The distribution of $\phi_{NBOH}$ resulting from the 35~ps AIMD trajectory does not satisfy the parity and $120^\circ$ periodicity expected from the symmetry of the surface. This is likely a sign of an insufficient sampling due to a high energy barrier preventing the hopping of the O-H bond above a B-N bond to another, equally probable energy minimum over the duration of the AIMD simulations. A fully equilibrated distribution should satisfy the above-mentioned parity and periodicity, which could be modeled as:
\begin{equation}
M(\phi_{NBOH}) = \frac{1}{2} \sum_{l=-1}^1 \left[ g(\phi_{NBOH} + 120.l) + g(\phi_{NBOH} + 2 \mu_k + 120.l) \right]
 \label{eq:modphi}
 \end{equation}
taking explicitly into account the identical role of the three B-N bonds, for $\phi_{NBOH}\in\{-120^\circ,0^\circ,120^\circ$. \}
This periodic and even model, also shown in Fig.~\ref{fig:phidistr} (with an arbitrary normalization to better visualize its dependence on the angle), is rather peaked around the positions above B-N bonds. For simplicity, in the classical MD simulations we treat the hydroxide groups as rigid (as the rest of the surfaces) and with one of the three most likely orientations.

An example of XYZ file of a hBN surface with one defect reads:
\small{
\begin{verbatim}
34
10.0181   8.676   0
B1  -1.2522725   0.723001446   0
B*  1.2522725    0.723001446   0.448818
B1  0           -1.445998554   0
B1  2.504545    -1.445998554   0
N  -1.2522725   -0.723001446   0
N   1.2522725   -0.723001446   0
N   0            1.445998554   0
N   2.504545     1.445998554   0
B1  3.7568175    0.723001446   0
B   6.2613625    0.723001446   0
B   5.00909     -1.445998554   0
B   7.513635    -1.445998554   0
N   3.7568175   -0.723001446   0
N   6.2613625   -0.723001446   0
N   5.00909      1.445998554   0
N   7.513635     1.445998554   0
B  -1.2522725    5.061001446   0
B   1.2522725    5.061001446   0
B1  0            2.892001446   0
B1  2.504545     2.892001446   0
N  -1.2522725    3.614998554   0
N   1.2522725    3.614998554   0
N   0            5.783998554   0
N   2.504545     5.783998554   0
B   3.7568175    5.061001446   0
B   6.2613625    5.061001446   0
B   5.00909      2.892001446   0
B   7.513635     2.892001446   0
N   3.7568175    3.614998554   0
N   6.2613625    3.614998554   0
N   5.00909      5.783998554   0
N   7.513635     5.783998554   0
\end{verbatim}
}
A suitable number of this modified hBN unit cell surface replaces the ones of regular hBN surface, according to the surface charge density. This choice is carried out randomly at the beginning of each trajectory, and so is the orientation of the OH bonds among the three possible choices (above the B$^*$-N bonds), with coordinates (for the B$^*$ indicated above):
\small{
\begin{verbatim}
O   1.2522725    0.723001446   2.005928
H   1.2522725   -0.1994633     2.3372737
\end{verbatim}
\begin{verbatim}
O   1.2522725    0.723001446   2.005928
H   0.45339379   1.18423252    2.3372737
\end{verbatim}
\begin{verbatim}
O   1.2522725    0.723001446   2.005928
H   2.05115121   1.18423252    2.3372737
\end{verbatim}
}

\section{Force field}
\label{sec:forcefield}

Interatomic interactions are described by 12-6 Lennard-Jones (LJ) pair potentials and Coulombic pairwise interaction : 
\begin{equation}
V_{ij} = 4 \epsilon_{ij} \left[ \left(\frac{\sigma_{ij}}{r_{ij}} \right)^{12} -  \left(\frac{\sigma_{ij}}{r_{ij}} \right)^{6}  \right] +  \frac{q_iq_j}{4 \pi \epsilon_0 r_{ij}}
\end{equation} 
where $r_{ij}$ is the interatomic distance between $i$ and $j$ atoms and $\epsilon_{ij}$ and $\sigma_{ij}$ are the LJ energy and diameter, respectively.  
Short-range interactions are computed using a cutoff $r_C$ = 12.0~\AA. Long-distance electrostatic interactions are computed with the particle-particle particle mesh method (P3M)~\cite{Hockney_1988} and a target accuracy of 1.10$^{-4}$ relative error on the forces.

\subsection{Lennard-Jones pair potentials}

The LJ parameters for both chemisorbed and physisorbed hydroxide groups are taken identical as that of the SPC/E model used for water~\cite{berendsenMissingTermEffective1987}. Those for potassium K$^+$ and chloride Cl$^-$ ions are taken from Ref.~\citenum{koneshanSolventStructureDynamics1998}, while that for graphite and hBN surfaces are taken from Refs.~\citenum{werderWaterCarbonInteraction2003} and~\citenum{wonWaterPermeationSubnanometer2007}, respectively. The Lorentz-Berthelot combination rules, are used for cross-interactions, \textit{i.e.}
\begin{equation}
\sigma_{ij} = \frac{\sigma_{ii} + \sigma_{jj}}{2} \quad {\rm{and}} \quad \epsilon_{ij} = \sqrt{\epsilon_{ii} \epsilon_{jj}}
\label{eq:LBrules}
\end{equation}
except between physisorbed hydroxide ions and the surfaces, for which the interaction strength is modified following:
\begin{equation}
\epsilon_{O_{HO^-}j_S} = 10 \sqrt{\epsilon_{O_{HO^-}O_{HO^-}} \epsilon_{j_Sj_S}}
\label{eq:LBexcept}
\end{equation}
where $j_S=$B for HBN surfaces and C for graphite surfaces. This parameter was determined empirically to ensure that hydroxide ions remain mostly in the vicinity of the surface.

\begin{table}
\begin{center}
\begin{tabular}{|c|c|c|c|}
\hline
Atom type $i$ & $\epsilon_{ii}$ (kcal.mol$^{-1}$)& $\sigma_{ii}$ (\AA) & $q_i$ ($e$)\\
\hline
\multicolumn{4}{|c|}{Electrolyte}\\
\hline 
Cl & 0.1000 & 4.401 & -1. \\ 
\hline 
K & 0.1000 & 3.331 & 1.\\ 
\hline 
O$_{H_2O}$ & 0.1554 & 3.169 & -0.8476 \\ 
\hline 
H$_{H_2O}$ & 0. & 0. & 0.4238  \\ 
\hline 
\multicolumn{4}{|c|}{Graphite surfaces}\\
\hline
C & 0.05649 & 3.211 & 0.  \\ 
\hline
\multicolumn{4}{|c|}{HBN surfaces}\\
\hline
B & 0.09484 & 3.451 & 0.37 \\
\hline
N & 0.1448 & 3.363 &  -0.37 \\
\hline
\multicolumn{4}{|c|}{Chemisorbed hydroxides}\\
\hline
B$^*$ & 0.09484 & 3.451 & 0.1298 \\
\hline
B$_1$ & 0.09484 & 3.451 & 0.314 \\
\hline
O$_{HO}$ & 0.1554 & 3.169 & -0.8476 \\ 
\hline 
H$_{HO}$ & 0. & 0. & 0.4238  \\ 
\hline 
\multicolumn{4}{|c|}{Physisorbed hydroxides}\\
\hline
O$_{HO^-}$ & 0.1554 & 3.169 & -1. \\ 
\hline 
H$_{HO^-}$ & 0. & 0. & 0.  \\ 
\hline 
\end{tabular}
\end{center}
\caption{Summary of pair potential parameters. The Lorentz-Berthelot combination rules (Eq.~\ref{eq:LBrules}) are used for all cross-terms except for the interaction of physisorbed hydroxide groups with the surface B or C atoms (see Eq.~\ref{eq:LBexcept}). 
}
\label{tab:LJparams}
\end{table}

\subsection{Coulombic interactions}

The partial charges $q_i$ are taken from the above-mentioned models of water and surface (see the references for the LJ parameters and Table~\ref{tab:LJparams}), except for the  chemisorbed and physisorbed hydroxide ions. For physisorbed ions, we follow the strategy of Bonthuis \textit{et al.} to assign a -1 charge on the oxygen atom and none to the H atom: Despite its simplicity (and resulting absence of dipole), the predictions of this model seem to be supported by experimental evidence~\cite{bonthuisOptimizationClassicalNonpolarizable2016}. Structure factors derived from small-angle x-ray scattering experiments shows very similar behaviors between (Na$^+$ + OH$^-$)$_{aq}$ and (Na$^+$ + F$^-$)$_{aq}$ solutions\cite{chenSolvationStructuresProtons2013}.

\begin{table}[ht!]
\begin{center}
\begin{tabular}{|c|c|c|c|}
\hline 
 & DFT (Vacuum) &  Force field  \\ 
\hline 
$Q_N$ & -0.35$^\dagger$  & -0.37\\ 
\hline 
$Q_B$ & 0.33$^\dagger$ &  0.37\\  
\hline
$Q_B$ $1^{st}$ neighbour & 0.33$^\dagger$ &  0.314 \\
\hline
$Q_{B^*}$ & 0.13 &  0.1298 \\ 
\hline 
$Q_{O^*}$ & -0.52 &  -0.8476 \\ 
\hline 
$Q_{H^*}$ & 0.19 & 0.4238 \\ 
\hline 
\end{tabular}
\end{center}
\label{tab:charges}
\caption{Mulliken charges extracted from DFT calculations in vacuum \cite{grosjeanVersatileElectrificationTwodimensional2019,grosjeanChemisorptionHydroxide2D2016} ($^\dagger$ averaged over all the atoms with the same atomic number in the cell) and partial charges used for the classical molecular dynamics simulations. }
\end{table}
 
Our choice of charges for chemisorbed hydroxide ions on hBN surfaces is based on the DFT calculations in vacuum of Grosjean \textit{et al.}~\cite{grosjeanVersatileElectrificationTwodimensional2019,grosjeanChemisorptionHydroxide2D2016}, summarized in Table~\ref{tab:charges}. However, some further considerations must be taken into account to design the classical charge distribution.
Firstly, in order to be consistent with the LJ interactions of the original force field for neutral hBN, we should also keep the corresponding partial charges, namely +0.37 and -0.37 for B and N atoms, respectively.
Secondly, the total charge of the defect (including the hydroxide group, the sp$^3$-hybridized B atom and the neighboring B atoms) must be equal to -1. Taking into account both constraints results in the charges summarized in Table~\ref{tab:charges}, where we use the same partial charges as the SPC/E water model for the surface O and H atoms, most of the excess charge resulting from DFT is equally shared between the 6 neighboring atoms B$_1$, while that of the hydroxide bearing atom B$^*$ is only slightly readjusted from the DFT one to strictly enforce the charge constraint.

\newpage
\section{Density profiles}
\label{sec:densprof}

 \begin{figure}[ht!]
\centering
        \includegraphics[width=0.8\columnwidth]{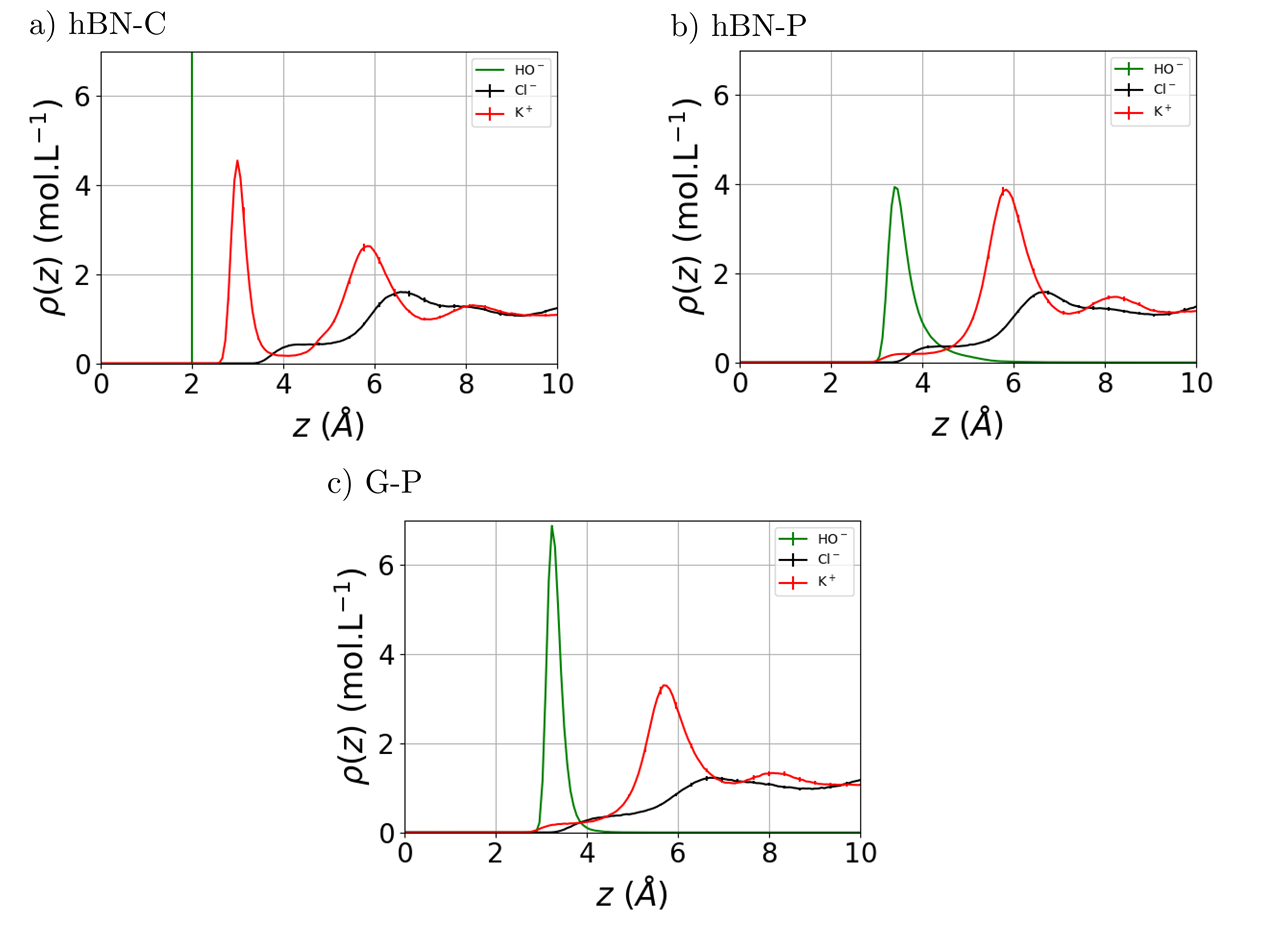}
    \caption{Density profiles of chloride (Cl$^-$), potassium (K$^+$) and oxygen atoms of hydroxide (HO$^-$) ions with a salt concentration of $c_s=1$~mol.L$^{-1}$ and titratable surface charge density  $\Sigma_{HO^-}=-0.02$~C.m$^{-2}$, for chemisorbed (a) and physisorbed (b) hydroxide ions on boron nitride as well as physisorbed hydroxide ions on graphite (c). In panel (a) the rigid chemisored hydroxide groups are fixed so that the density of hydroxide ions appears as a vertical line. }
    \label{fig:densprof}
\end{figure}

In order to illustrate the ability of the above force field to feature the behavior of hydroxide ions expected from AIMD simulations, Fig.~\ref{fig:densprof} reports the density profiles of all ionic species (from which the charge density profiles, see Fig.~6 of the main text, are computed) for a salt concentration of $c_s=1$~mol.L$^{-1}$ and titratable surface charge density  $\Sigma_{HO^-}=-0.02$~C.m$^{-2}$. 
In the chemisorbed case (a), the hydroxide groups are fixed on the surface by construction. In the physisorbed ones (panel b for hBN and panel c for graphite), the tailored force field, with a strong Lennard-Jones interaction between the oxygen of hydroxide groups and the surface results in the expected physisorption characterized by a single peak of the density profile near the surface (for completeness, we also show the density profiles in the case of hBN with chemisorbed hydroxide ions, which appears for the latter in a vertical line in panel a).